\documentclass[12pt]{article}
\usepackage{amsmath,amssymb,epsfig}
\usepackage{graphicx}

\usepackage{amsmath,amssymb}
\usepackage{graphicx}
\usepackage{amsmath,amssymb}
\usepackage{graphicx}

\newcommand{\be}{\begin{equation}}
\newcommand{\ee}{\end{equation}}
\newcommand{\bea}{\begin{eqnarray}}
\newcommand{\eea}{\end{eqnarray}}
\newcommand{\ba}{\begin{array}}
\newcommand{\ea}{\end{array}}

\def \nn {\nonumber}
\newcommand{\eq}[1]{(\ref{#1})}

\renewcommand{\title}[1]{\vbox{\center\LARGE{#1}}\vspace{5mm}}
\renewcommand{\author}[1]{\vbox{\center#1}\vspace{5mm}}

\newcommand{\bR}{\mathbb{R} }
\newcommand{\bZ}{\mathbb{Z} }
\newcommand{\bC}{\mathbb{C} }

\newcommand{\vap}{\varphi}

\numberwithin{equation}{section}
\allowdisplaybreaks

\begin{document}

\begin{titlepage}
\begin{flushright}
\end{flushright}

\vfill

\begin{center}
{\large \bf
Holographic Anyons in the ABJM Theory
\vfill
{
Shoichi Kawamoto\footnote{\tt kawamoto@ntnu.edu.tw}
}
and
{
Feng-Li Lin\footnote{\tt linfengli@phy.ntnu.edu.tw}
}

\bigskip
{\it
Department of Physics,
National Taiwan Normal University, \\
Taipei, Taiwan 11677
}}

\end{center}

\vfill
\begin{abstract}
We consider the holographic  anyons in the ABJM theory from three
different aspects of AdS/CFT correspondence. First, we identify the
holographic anyons by using the field equations of supergravity,
including the Chern-Simons terms of the probe branes.
We find that the composite of D$p$-branes wrapped over $\mathbf{CP}^3$
with the worldvolume magnetic fields can be the anyons.
Next, we discuss the possible candidates of the dual anyonic operators
on the CFT side, and find the agreement of
their anyonic phases with the supergravity analysis.
Finally, we try to construct the brane profile for the
holographic anyons by solving the equations of motion and Killing
spinor equations for the embedding profile of the wrapped branes.
As a by product, we find a BPS spiky brane for the dual baryons in the
ABJM theory.
\end{abstract}
\vfill

\end{titlepage}

\setcounter{footnote}{0}

\section{Introduction}

Anyons proposed in
\cite{Leinaas:1977fm,Wilczek:1981du,Wilczek:1982wy} are the point
particles obeying the fractional statistics, and  they exist in
$2+1$ dimensions because the linking number in $2+1$ dimensions is
well-defined.
When the positions of two anyons are interchanged,
the wavefunction of the system will get a fractional phase.
Moreover, these anyonic
particles can be described by $U(1)$ Chern-Simons effective
field theory \cite{Wilczek:1983cy}.
Namely, via the coupling to the Chern-Simons term
the electrons are endowed with a fictitious magnetic
flux, which will induce Aharonov-Bohm (AB) phase
when one is going around another.
Since the exchange of two particles
is considered as the half-winding,
this AB phase is responsible for the fractional
statistics.
One example of the above construction is the effective
field theory of the quasiparticles for
the fractional quantum Hall fluids
\cite{Zee95}.
The generalization to use D-branes/noncommutative
Chern-Simons for describing quantum Hall fluids can be seen in
\cite{Brodie:2000yz}, and the related holographic construction was
done recently in \cite{Fujita:2009kw}.

Since the anyon is usually thought of as a quasiparticle in a strongly
coupled system, and cannot be seen in the perturbative approach. It
is then interesting to see if one can construct anyons from the
holographic dual of some strongly coupled systems so that anyons can
be realized as D-brane configurations.
Motivated by the relation between anyons and the $(2+1)$-dimensional
Chern-Simons theory, it seems that the recently constructed
Aharony-Bergman-Jafferis-Maldacena (ABJM) theory
 \cite{Aharony:2008ug} is a good starting point for our purpose,
since the ABJM theory is given as $\mathcal{N}=6$ superconformal
Chern-Simons-matter theory in $2+1$ dimensions and also its gravity
dual of type IIA supergravity in $AdS_4 \times \mathbf{CP}^3$
background is known.
Naively, the holographic anyon we are seeking for should be quite
different from the one in the usual $U(1)$ Chern-Simons effective
field theory.
The main reason is that
the latter is a quasiparticle consisting of electrons
and is thus not gauge invariant, and therefore would not be observed
in the bulk gravity side into which only the gauge invariant states
are mapped.
However, as we will see the
holographic anyons are indeed not gauge invariant states in the dual
field theory, but still can be observed.

We can generalize the case with $U(1)$ Chern-Simons to the non-abelian
one by introducing the 't Hooft disorder operators.
They are
defined as the large gauge transformation along a given contour,
and also known as 't Hooft loop.
If the
theory contains no charged matter under the center of the gauge group,
the 't Hooft disorder operator is local,
that is, any field in the action cannot detect the presence of the 't
Hooft operator. 
However, as shown in \cite{Itzhaki:2002rc}, in the presence of the
Chern-Simons term the 't Hooft
operators can detect each other and thus behave like
anyons.
In the ABJM case, the 't Hooft operators are attached to the chiral
primary operators that are dual to the D0-brane and D4-branes wrapping
on a cycle inside $\mathbf{CP}^3$, and  makes these operators gauge
invariant.
Therefore, such gauge invariant states are by definition local under
large gauge transformation, and cannot be the anyons.

A way out is to consider the wrapped D2 and D6-branes over
$\mathbf{CP}^3$.
There should be fundamental strings stretching between the branes
and the boundary due to the charge conservation on the world volume.
They are dual to the baryon
vertex \cite{Aharony:2008gk} in the field theory side,
and are similar to the wrapped D5-brane in the case of $AdS_5\times S^5$
 \cite{Witten:1998xy}-\cite{Brandhuber:1998xy}.
Moreover, these baryons are not the gauge invariant states due to the
fundamental strings stretching to the boundary.
They will therefore pick up a
fractional phase under the action of 't Hooft loop.
It then implies
that we can make anyons in the ABJM theory by considering the bound
states of baryons and 't Hooft disorder operators.
Indeed, we find that the
holographic anyons are what we call the dressed baryons, namely, the bound state
 of baryons, chiral primaries and 't Hooft disorder operators.

With the above consideration in mind, one can directly look for the
holographic anyons by studying the supergravity equations of motion
with probe brane sources.
Indeed, in \cite{Hartnoll:2006zb} Hartnoll
has used this approach to construct anyonic strings and membranes in
various AdS spaces.
The supergravity Chern-Simons terms which couple the background
fluxes are
responsible for the resultant fractional AB phases when winding one
probe brane around the other.
In the ABJM theory, we can also
construct this kind of anyonic D0-brane and F-string pair, as well as
the D4-F1 pair.
They are
holographic anyons, but their anyonic phases are $1/N$ suppressed
in the 't Hooft limit as in \cite{Hartnoll:2006zb}.

The holographic anyons we will construct are made of baryonic spiky
branes and magnetic fluxes introduced on their world-volumes.
On these baryonic branes we need to attach either $k$ or $N$
fundamental strings to satisfy the charge conservation condition.
Since $k$ and $N$ will be of order $N$ quantities, then the anyonic
 phases gained by the set of strings are no longer suppressed in the
 large $N$ limit.
Moreover, the magnetic fields on the
D-brane worldvolume can be thought of as dissolved D-branes,  these
holographic anyons are in fact bound states of particle-like branes wrapped
on the internal $\mathbf{CP}^3$.
Obviously, these holographic anyons
should be the anyonic dressed baryons discussed above.
We find that these anyons have the anyonic phases
proportional to  either the 't Hooft coupling or its inverse
 \footnote{In contrast,
it is interesting to note that the fractional phase for the edge
states of FQHE from D-brane construction in \cite{Fujita:2009kw} is
proportional to 't Hooft coupling.}, which will not be suppressed in
the 't Hooft limit.

More precisely, from the supergravity analysis we find that the
anyonic phase arises either from winding the spiky D2 with $k$
fundamental strings around the D0 (including the dissolved ones on
higher wrapped branes), or from winding the spiky D6 with $N$
fundamental strings (with the opposite orientation to the spiky D2's)
around the the wrapped D4. Furthermore, we find the agreement of these
anyonic phases with the ones from the field theory analysis, where the
non-perturbative effect due to 't Hooft loop is involved.

We also explore these particle-like branes
and their bound states from the open string picture by solving
the worldvolume's equations of motion and the Killing spinor equations
for supersymmetric embedding of wrapped D-branes. As a by-product, we
find the BPS D2 spiky baryonic-branes in the ABJM theory. In terms
of the form of the solution these spiky solutions are quite
nontrivial.
However,  our holographic anyons are not BPS states.

We organize the paper as follows. In the next section we will review
Hartnoll's idea \cite{Hartnoll:2006zb}  on constructing the anyonic
branes with new examples in the ABJM theory, and also set up the
notations. In the section 3 using field equations of supergravity we
show that some wrapped branes with magnetic fluxes can be realized as
the holographic anyons in the ABJM theory.
We find that the phase is essentially given by strings from D2-baryon
 going around D0-branes, or D6-baryon around wrapped D4-branes.
In the section 4 we discuss the field theory interpretation of the
 holographic anyons found in the previous section.
Moreover, we find that the anyonic phases from
 field theory analysis agree with the ones found in the supergravity side.
In the section 5 we solve the equations of motion and the Killing
spinor equations for the wrapped D-branes of holographic anyons. We
conclude our paper in the section 6. Some minor details, convention
setup and useful formulas are put in two Appendices A and B, and the
details of our trial for solving BPS D4 and spiky D6 wrapped branes in
Appendix C.

\section{Anyonic pair in ABJM}

The anyonic branes were first considered by Hartnoll in
\cite{Hartnoll:2006zb} where he used field equations of supergravity
to show that some pair of branes will pick up a fractional AB phase
when one brane transverses the other, and thus are anyonic.  He
considered the examples of F-D strings in $AdS_5\times S^5$ and
membranes in $AdS_7 \times S^4$ based on the nontrivial topology of
configuration space in higher dimensions such as $H_2(\bR^4\setminus
\bR)=H_3(\bR^6\setminus \bR^2)=\bZ$.  Instead of reciting his
examples, we consider an anyonic pair of branes in the ABJM
theory,
and at the same time set up the notations.
As we shall see, the anyonic pair is the D0-brane and F1-string based
on the nontrivial topology of configuration space $H_1(\bR^3
  \setminus \bR)=\bZ$ for D0 going around F1 and $H_2(\bR^3 \setminus
  \{0\})=\bZ$ for F1 surrounding D0.

  We start with the relevant part of the IIA action in the Einstein frame
  with also the source terms due to the presence of various strings
  and branes \cite{Myers:1999ps}
\bea
\label{IIA_SUGRA_action}
L=&-&{1\over 4\kappa^2_{10}}\int \left( e^{-\Phi} H_3 \wedge * H_3 +e^{3\Phi\over 2} F_2\wedge *F_2+e^{\Phi/2} \tilde{F}_4\wedge * \tilde{F}_4+B_2 \wedge F_4\wedge F_4 \right)\nn \\
&-& \mu_0 \int_{D0} C_1 -T_{F1} \int_{F1} B_2 -\mu_2 \int_{D2}\left(C_3+C_1\wedge {\tilde B}_2\right)
\nn\\
&-&\mu_4 \int_{D4} \left(C_5+C_3\wedge {\tilde B}_2+{1\over 2} C_1\wedge {\tilde B}_2\wedge {\tilde B}_2\right) \nn\\
&-& \mu_6 \int_{D6} \left(C_7+C_5\wedge {\tilde B}_2 +{1\over 2} C_3\wedge {\tilde B}_2\wedge {\tilde B}_2+{1\over 6} C_1\wedge {\tilde B}_2\wedge {\tilde B}_2\wedge {\tilde B}_2\right)\;.
\eea
where $\tilde{F}_4=F_4-C_1\wedge H_3$, $2\kappa^2_{10}=(2\pi)^7
\ell_s^8$,  $T_{F1}=\mu_1$, and $\mu^2_p={\pi \over \kappa^2_{10}}
(2\pi\ell_s)^{2(3-p)}$ so that $2\kappa^2_{10} \mu_p=(2\pi
\ell_s)^{7-p}$, and the integrations  $\int_{Dp}$ or $\int_{F1}$ are
integrated over the worldvolume of the corresponding source
branes. Moreover, ${\tilde B}_2=B_2+2\pi \ell_s^2 dA_1$ with $dA_1$
the magnetic fluxes being turned on in the D-brane
worldvolume. Hereafter, we will set $\ell_s=1$ for simplicity.

  From the action, we can derive the field equations for $C_3$, $B_2$
  and $C_1$ as follows:
\be\label{C3}
d*(e^{\Phi/2}\tilde{F}_4)=-H_3\wedge F_4 - 2\kappa^2_{10}\mu_2 \delta^7(x_{D2})-2\kappa^2_{10}\mu_4 {\tilde B}_2 \delta^5(x_{D4})-\kappa^2_{10}\mu_6 {\tilde B}_2\wedge {\tilde B}_2 \delta^3(x_{D6})\;,
\ee
\bea
& & d*(e^{-\Phi}H_3)+F_2\wedge *(e^{\Phi/2}\tilde{F}_4)+C_1\wedge H_3\wedge F_4-{1\over 2} F_4\wedge F_4 -2\kappa^2_{10}T_{F1}\delta^8(x_{F1})\nn\\ \label{B2}
&& -2\kappa^2_{10}\mu_4 C_3\delta^5(x_{D4})-2\kappa^2_{10}\mu_6 (C_5+C_3\wedge {\tilde B}_2)\delta^3(x_{D6})=0\;,
\eea
and
\bea
& & d*(e^{3\Phi/2}F_2)-H_3\wedge *(e^{\Phi/2}\tilde{F}_4)+2\kappa^2_{10}\mu_0\delta^9(x_{D0})+2\kappa^2_{10}\mu_2 {\tilde B}_2 \delta^7(x_{D2})
\nn \\ \label{C1}
&& + \kappa^2_{10}\mu_4 {\tilde B}_2\wedge {\tilde B}_2 \delta^5(x_{D4})+{1\over 3} \kappa^2_{10}\mu_6 {\tilde B}_2 \wedge {\tilde B}_2 \wedge {\tilde B}_2 \delta^3(x_{D6})=0\;,
\eea
where $\delta^{9-p}(x_p)$ is the Poincar\'e dual $(9-p)$-form to the
worldvolume of Dp-brane as defined by $\int_{Dp} C_{p+1}=\int C_{p+1}
\wedge \delta^{9-p}(x_p)$ with the second integral over the whole
spacetime.
Note that we arrive \eq{B2} by using \eq{C3} so that the terms like
$C_1\wedge \tilde{B}_2 \delta^5(x_{D4})$ and $C_1\wedge \tilde{B}_2
\wedge \tilde{B}_2 \delta^3(x_{D6})$ are canceled out.

 We will consider the ABJM background \cite{Aharony:2008ug} as
 follows (see Appendix \ref{appendix2} for more detailed
  expressions.)
\be
ds_E^2=e^{-\Phi^{(0)}/2}{R^3\over k}({1\over 4} ds^2_{AdS_4}+ ds^2_{CP^3})
\ee
and
\be
e^{2\Phi^{(0)}}={R^3\over k^3}\;,
\qquad F_4^{(0)}=dC_3^{(0)}={3\over  8}R^3  \hat{\epsilon}_4\,
\qquad F_2^{(0)}=dC_1^{(0)} = {k} J
\ee
where the superscript $(0)$ denotes that they are the background
values, $\hat{\epsilon}_4$ is the volume element of unit $AdS_4$, and
$J$ is proportional to the K\"ahler form of $\mathbf{CP}^3$.

From this, we have
\be
*(e^{\Phi^{(0)}/2} F_4^{(0)})={6\over k}R^6  \hat{\epsilon}_6
\ee
where $\hat{\epsilon}_6$ is the volume element of unit
$\mathbf{CP}^3$. Moreover, the volume of unit $\mathbf{CP}^3$ is the
same as unit $6$-sphere's, i.e., $vol_{\mathbf{CP}^3}={\pi^3\over 6}$.

Now, let us consider adding a source D0 particle in the ABJM background and
then transversing a probe F1 string around it. From \eq{C1}, at
the linear order in the fluctuations we have
\be
H_3\wedge \hat{\epsilon}_6 ={k \over 6 R^6} \left(
  (2\pi)^7 \delta^9(x_{D0})+ e^{3\Phi^{(0)}/2} d*F_2
+\frac{3}{2} d*(e^{3\Phi^{(0)}/2} \Phi F_2^{(0)}) \right)
\ee
where $H_3$, $F_2$ and $\Phi$ are linear perturbations on top of the
ABJM background due to the presence of the D0 source.

Then, there is an asymptotic anyonic phase picked up by the partition
 function of the fundamental string probe
 after transversing around the D0 as follows
\bea
\Delta \phi_{F1D0}&=&-T_{F1}\int_{\partial \Sigma} B_2\nn\\
&=&-T_{F1} {1\over vol_{\mathbf{CP}^3}} \int_{\partial \Sigma\times \mathbf{CP}^3} B_2 \wedge \hat{\epsilon}_6 \qquad\qquad\qquad  [\mbox{as}\;\; r_{min}\longrightarrow \infty]
\nn\\
&=&-T_{F1} {1\over vol_{\mathbf{CP}^3}} \int_{\Sigma\times \mathbf{CP}^3} H_3 \wedge \hat{\epsilon}_6\nn\\
&=& -T_{F1} {1\over vol_{\mathbf{CP}^3}} {k\over 6 R^6 }
\int_{\Sigma\times \mathbf{CP}^3}\left( (2\pi)^7
  \delta^9(x_{D0})+ e^{3\Phi/2} d*F_2
+\frac{3}{2} d*(e^{3\Phi^{(0)}/2} \Phi F_2^{(0)}) \right)\nn\\
&=& -{2\pi \over N} - {2\pi \over N} {1\over (2\pi )^7} e^{3\Phi^{(0)}/2}  \int_{\partial \Sigma\times \mathbf{CP}^3} *F_2. \label{f1d0phase}
\eea
where $\partial \Sigma$ is the worldvolume swept by the probe fundamental
string. In the above the second step is justified by assuming the
minimal distance $r_{min}$ between F1 and D0 are large enough so that
$B_2$ sourced by D0 is independent of the $\mathbf{CP}^3$ coordinates,
i.e., suppressing the higher harmonics. Moreover, in the third step we
have used the Stokes' theorem, for its validity we need to close up
the swept surface $\partial \Sigma$ with the D0 being enclosed. However, if the fluctuations
$F_2$ and $H_3$ are suppressed in the large $r$ limit, we can just fix
the IR end point of F1, and move the UV one to sweep out a cone-like surface $\partial \Sigma$
in the bulk as shown in Fig. \ref{F1-D0}. Indeed, as shown in the
Appendix \ref{appendix1}, both $F_2$ and $H_3$ are massive and will be
suppressed in the large $r$ limit. Moreover, this also implies that
the dynamical part of the above phase, i.e., $\int *F_2$, can be
neglected if the separation between the source and probe is large
enough.
Finally, from the fourth line to the fifth line, we have used the fact
that since $F_2^{(0)}$ is the 2-form inside the
$\mathbf{CP}^3$, $d * (\Phi F_2^{(0)})$ is a 9-form schematically
written as $\hat\epsilon_4 \wedge f_5$ where $f_5$ is a 5-form
inside the $\mathbf{CP}^3$, and thus it trivially vanishes in the integral.

\begin{figure}[htb]
\begin{center}
\includegraphics[width=20em,clip]{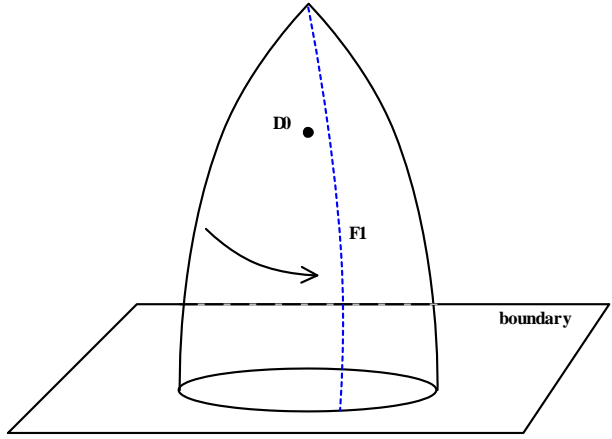}
\end{center}
\caption{The world sheet of fundamental string surrounding D0-brane.}
\label{F1-D0}
\end{figure}

Similarly, we consider putting a source F1 string in the ABJM
background and take a probe D0 particle to transverse it.
 From
 \eq{B2}, at linear order in
the fluctuations we have
\begin{align}
  F_2 \wedge \hat{\epsilon}_6 =& {k\over 6 R^6 } \left(
    (2\pi)^6\delta^8(x_{F1})-e^{-\Phi^{(0)}} d* H_3
-e^{\Phi^{(0)}/2} F_2^{(0)} \wedge * ( \tilde F_4)
- P_1 \right)\;.
\\
\label{def_P1}
P_1 =& \frac{1}{2}F_2^{(0)} \wedge * (e^{\Phi^{(0)}/2}\Phi \tilde F_4^{(0)})
+C_1^{(0)}\wedge H_3 \wedge F_4^{(0)}
-F_4^{(0)} \wedge F_4 \,,
\end{align}
where note that the first term in $P_1$ vanishes trivially.
Then, there is an asymptotic anyonic phase picked up by the probe D0 for its transverse motion around F1 as follows
\bea
\Delta \phi_{D0F1}&=&-\mu_0\int_{\partial \Sigma} C_1\nn\\
&=&-\mu_0 {1\over vol_{\mathbf{CP}^3}} \int_{\Sigma\times \mathbf{CP}^3} F_2 \wedge \hat{\epsilon}_6 \qquad\qquad\qquad  [\mbox{as}\;\; r_{min}\longrightarrow \infty]\nn\\
&=& -\mu_0 {1\over vol_{\mathbf{CP}^3}} {k\over 6 R^6 }
\int_{\Sigma\times \mathbf{CP}^3}\left(
  (2\pi)^6\delta^8(x_{F1})-e^{-\Phi^{(0)}} d* H_3
+e^{\Phi^{(0)}/2} F_2^{(0)} \wedge * ( \tilde F_4)
+P_1 \right)\nn\\
&=& -{2\pi \over N} + {2\pi \over N} {1\over (2\pi )^6}
e^{-\Phi^{(0)}}  \int_{\partial \Sigma\times \mathbf{CP}^3} *H_3,
\eea
where we have used again the fact that $P_1$ vanishes trivially inside
the integral due to its form structure.
From the third line to the fourth line, we have set $\tilde F_4=0$
\footnote{%
We do not have any dynamical reason to drop this term, however, the
phase obtained here should be the same as the one in \eq{f1d0phase}
because of their symmetrical situations.
Then we simply assume $\tilde F_4=0$ which is consistent with the
equations of motion.}.
This is allowed since from the equations of motion \eq{C3} and \eq{C1}
we have
\begin{align}
0=&  d*(e^{\Phi^{(0)}/2} \tilde F_4)
+\frac{1}{2}  d*(e^{\Phi^{(0)}/2}\Phi  F_4^{(0)})
+H_3 \wedge F_4^{(0)} \,,
\nn\\
0=& d*(e^{3\Phi^{(0)}/2} F_2)
+\frac{3}{2}  d*(e^{3\Phi^{(0)}/2}\Phi  F_2^{(0)})
-e^{\Phi^{(0)}/2} H_3 \wedge *(F_4^{(0)}) \,, \nn
\end{align}
and these can be solved by $\tilde F_4 =0$, $\Phi \neq 0$ and $H \neq
0$.
Finally, in the last line the dynamical part associated with the integral $\int
*H_3$ can be neglected as
before.


The phase we have obtained here is proportional to $1/N$
and then will vanish in the large $N$ limit.
In the next section, we will introduce spiky D-branes
that have a straightforward interpretation as baryons,
which also allow us to consider the dual field theory
counterpart.
As we will see, the anyonic phase for the dressed baryon will be
proportional to $1/\lambda$ which will not be suppressed in the large
$N$ limit.

    Moreover, by considering the field equations for D4 source, we find that
F1 and D4 form the anyonic pair with the anyon phase proportional to $1/k$,
which is suppressed in the large $k$ limit. The detailed analysis  is similar to
what will be done in arriving \eq{anyonpairf1d4}, thus we will omit it here.

In summary: The above anyonic pair is similar to the ones considered
in \cite{Hartnoll:2006zb} for the IIB and M-theory cases. However,
in all these cases the anyoic phase goes to zero in the large $N$ or $k$
limit, hence it plays no role in the holographic consideration when
the large $N$ or $k$ limit is taken. In the next section we will consider
the case with nontrivial anyonic phase even when the large $N$ or $k$ limit
is taken but $k/N$ is fixed.
Especially they can be understood as the anyonic
particles on the boundary dual Chern-Simons theory.

\section{Holographic anyons in ABJM}

  Based on the similar supergravity approach, we now show that a
  slight generalization of the particle-like branes proposed in
  \cite{Aharony:2008ug} are in fact the holographic anyons with the
  anyonic phases surviving even in the large $N$ and $k$ limit.
Since we want these holographic anyons to be particle-like on
the AdS boundary, so it should be the particle-like branes as
discussed in \cite{Aharony:2008ug}.
We will see that there is an extra
ingredient to have nontrivial anyonic phase, which is to introduce
dressed baryonic branes.

   These holographic anyons are constructed as following. Wrapping
   $n_6$ D6-branes over $\mathbf{CP}^3$, $n_4$ D4-branes on a
   $\mathbf{CP}^2$(or $\mathbf{CP}^1 \times \mathbf{CP}^1$) cycle of
   $\mathbf{CP}^3$ and $n_2$ D2-brane on a $\mathbf{CP}^1$ cycle of
   $\mathbf{CP}^3$. These D-branes look as particles in the $AdS_4$
   located at some radial distance, which can be combined with $n_0$
   D0-branes to form a composite particle.    Besides, we will also
   turn on the magnetic fluxes denoted by $dA_1$ on the worldvolumes
of D6, D4 and D2 wrapping over the cycles on $\mathbf{CP}^3$ such
that
\be\label{d6B}
\frac{1}{6}\int_{\mathbf{CP}^3} {dA_1 \over 2\pi} \wedge {dA_1 \over 2\pi} \wedge {dA_1 \over 2\pi} = m_6,
\ee
\be\label{d4B}
\frac{1}{2}
\int_{\mathbf{CP}^2} {dA_1 \over 2\pi} \wedge {dA_1 \over 2\pi}= m_4,
\ee
and
\be\label{d2B}
\int_{\mathbf{CP}^1} {dA_1 \over 2\pi} =  m_2
\ee
where $m_i$'s are integers, and can be understood as relating to the
number of dissolved D0 branes
 \footnote{From the Chern-Simons terms of probe D6 and D4, the magnetic fields can also induce 
 D4 (on probe D6) and D2 (on probe D6 and D4) charges. However, unlike the induced D0's, these charges will not contribute to the anyonic phases considered here.}
and the associated linking numbers.
Note that though we have used the same $A_1$ for all the cases, they
are all different gauge fields on different branes.

Moreover, as pointed out in \cite{Aharony:2008ug}, the D6-brane
worldvolume Wess-Zumino(WZ) coupling $\int_{D6} A_1\wedge *F_4$
implies that there are $n_6 N$ fundamental strings ending on it. Or, the
D6-brane has a spiky shape, similar to the D5 baryonic branes
considered in
\cite{spike-solutions,Gibbons:1997xz,Witten:1998xy,Imamura:1998gk}.
Similarly, the WZ coupling $\int_{D2} A_1\wedge F_2$ on D2-brane
worldvolume implies that there are $n_2 k$ F-strings ending on
it. However, the orientations of the above two kind of F-strings are
opposite so that the net number of fundamental strings is $n_6 N- n_2 k$. These
fundamental strings will stretch like a spike from the wrapped branes and end
on the $AdS_4$ boundary, which looks as a composite particle on the
boundary, namely, the baryon \footnote{As for baryonic D5-branes, the
spiky brane configuration is BPS but the one with fundamental strings ending
on the brane is not.}. Since these baryons would be dressed by the chiral
operators dual to the induced D0- and D4-brane charges, we will call
them the dressed baryons. We will show that these dressed baryons are in fact
anyons.

If we turn on such a spiky particle-like brane as a source in the
ABJM background, then from \eq{C1}, to the linear order we
have\footnote{%
\eq{C3} and \eq{B2} would give the contributions to the anyonic phase from
D2 and fundamental string charges respectively, but it turns out that they will not
have any effect on fundamental string going around the particle-like branes.
It would be because these wrapping branes do not carry the net charges,
only multipoles, and then their effects are neglected in our approximation.}
\bea
&& H_3\wedge \hat{\epsilon}_6 ={k \over 6 R^6}
\left( e^{3\Phi^{(0)}/2} d*F_2 + \frac{3}{2}e^{3\Phi^{(0)}/2} d*(\Phi
  F^{(0)}_2)  \right)
+ vol_{\mathbf{CP}^3} \times {1\over T_{F1}} \times \frac{2\pi}{N} \times\nn \\
&&  \left(n_0 \delta^9(x_{D0})+{dA_1\over 2\pi} n_2 \delta^7(x_{D2}) +{1\over 2}  {dA_1\over 2\pi} \wedge {dA_1\over 2\pi} n_4 \delta^5(x_{D4}) +{1\over 6} {dA_1 \over 2\pi} \wedge {dA_1 \over 2\pi} \wedge {dA_1 \over 2\pi} n_6 \delta^3(x_{D6})  \right) \nn
\eea
where $H_3$, $F_2$ and $\Phi$ are linear perturbations on top of the ABJM
background due to the presence of the particle-like brane source.
 Then, the asymptotic anyonic phase picked up by  winding the F1 around the particle-like brane is
\bea
\Delta \phi_{F1D0} &=&-T_{F1}\int_{\partial \Sigma} B_2\nn\\
&=& -T_{F1} {1\over vol_{\mathbf{CP}^3}} \int_{\Sigma\times \mathbf{CP}^3} H_3 \wedge \hat{\epsilon}_6 \qquad\qquad\qquad  [\mbox{as}\;\; r_{min}\longrightarrow \infty] \nn\\ \label{hphase}
&=& -{2\pi \over N}(n_0+n_2m_2+n_4m_4+n_6m_6) -
  {2\pi \over N} {1\over (2\pi )^7} e^{3\Phi^{(0)}/2}  \int_{\partial
  \Sigma\times \mathbf{CP}^3} *F_2 \,,
\eea
and again $d*(\Phi  F^{(0)}_2)$ part vanishes trivially in the integral.
The second term in \eq{hphase} can be neglected as usual due to the
massive nature of $F_2$.  Note also that the IR end point is fixed
when we sweep the F-string, and the swept surface is closed up at IR end.

Now we consider two spiky particle-like branes with quantum number
$(n_i,m_i,r)$ and $(n'_i,m'_i,r')$ for $i=0,2,4,6$ except there is no
$m_0$ and $m'_0$.
$r$ and $r'$ are their radial locations respectively.
Consider that $r'\gg r$.
Once we exchange these two baryons on the
boundary, it will induce an equivalent anyonic phase by the following
two viewpoints.

The first is that
the $(n_i,m_i,r)$ particle is enclosed by the swept F-strings attached to the
$(n_i',m_i',r')$ particle, i.e.,
\bea
\Delta \phi_{D0D2}&=&{2\pi \over N}(n'_2 k-n'_6 N)(n_0+n_2m_2+n_4m_4+n_6m_6) \nn \\ \label{abphase1}
&=& 2\pi ({n'_2\over \lambda} - n'_6) (n_0+n_2m_2+n_4m_4+ n_6m_6)  \; \mbox{mod} \; 2\pi,
\nn\\
&=& 2\pi \frac{n'_2 N_0}{\lambda} \,,
\eea
where $\lambda=N/k$ is the 't Hooft coupling, which is fixed in the ABJM
theory
when taking large $N$ and $k$ limit
and $N_0=n_0+n_2m_2+n_4m_4+ n_6m_6$ is the total D0 brane charge
carried by $(n_i,m_i,r)$ particle-like brane.
The second is that
the $(n_i,m_i,r)$ particle is moving around the F1 strings
attached to the $(n_i',m_i',r')$ particle.
Note that we regard $r'$ as virtually infinity, since otherwise the
surface $\Sigma$ spanned by the orbit of D0 is not well defined.
Now the linearized equations of motion leads
\footnote{We again set $\tilde F_4=0$ for the same reason as in the footnote 2.}
\begin{align}
F_2 \wedge \hat{\epsilon}_6 = {k\over 6 R^6 }&  \left(
  (2\pi)^6 (n'_6 N - n'_2 k)\delta^8(x_{F1})
+(2\pi)^3 n_4' C^{(0)}_3 \delta^5(x_{D_4})
+2\pi n_6' C^{(0)}_3 \wedge  \tilde{B}_2 \delta^3(x_{D6})
\right.\nn\\&\left.
-e^{-\Phi^{(0)}} d* H_3
-F_2^{(0)} \wedge * (e^{\Phi^{(0)}/2} \tilde F_4)
-P_1
 \right)\;,
\end{align}
where $P_1$ is given in \eq{def_P1}.
In this expression only the first term will contribute to the
phase as
\begin{align}
\Delta \phi_{D0D2}=&
-\int_{\partial \Sigma} C_1 \times \left(\mu_0+\mu_2
  \int_{\mathbf{CP}^1} \tilde{B}_2 +{1\over 2} \mu_4
  \int_{\mathbf{CP}^2} \tilde{B}_2 \wedge \tilde{B}_2 +{1\over 6}
  \mu_6 \int_{\mathbf{CP}^3} \tilde{B}_2 \wedge \tilde{B}_2 \wedge
  \tilde{B}_2 \right)
\nn\\=&
- 2\pi \frac{n'_2 N_0}{\lambda} \,.
\end{align}
In this result, the minus sign of the phase
corresponds to the fact that now the
particle-like branes go around the F-strings in the opposite
direction.
Note that this anyonic phase survives in the large $N$ limit
and is coupling dependent.
Moreover the phase is basically given by winding the F-strings of the spiky
wrapped D2-brane around the D0-branes, which includes the ones
induced (or dissolved) on higher dimensional branes' world-volumes.

Now we move to the anyonic phase regarding D4-brane source.
To see this type of
holographic anyons, we can inspect the field equation of the 6-form
flux, and at linearized level it is
\begin{align}
   H_3 \wedge F^{(0)}_2 =&
 d * \left(e^{-\Phi^{(0)}/2} F_6 \right)
-\frac{1}{2} d * \left(e^{-\Phi^{(0)/2}} \Phi F^{(0)}_6 \right)
+2\kappa_{10}^2 \mu_4 \delta^5 (x_{D4}) \,,
\end{align}
which can be obtained as the Bianchi identity of 2-form field strength
\cite{bi}.
With this, we now consider the
winding of a F1 around a wrapped D4 over a 4-cycle
inside $\mathbf{CP}^3$.



We have the 2-form $J=d\omega$ which
is proportional to the K\"aler form of $\mathbf{CP}^3$.
We here consider D4-brane wrapping on a four cycle which is dual
to the 4-form $J \wedge J$.
Since $J \wedge J \wedge J = -48 \hat\epsilon_6$,
the Poincar\'e dual inside $\mathbf{CP}^3$ to the world volume of D4 is
given by $\delta^2 = -J/(2\pi)$.
Thus, the associated AB phase is
\begin{align}
  \Delta \phi_{F1D4} =&
-T_{F1} \int_{\partial \Sigma} B_2
\nn\\=&
-T_{F1}\frac{1}{-48 vol_{\mathbf{CP^3}}}
 \int_{\Sigma \times \mathbf{CP}^3}
H_3 \wedge J \wedge J \wedge J
\nn\\=&
T_{F1}\frac{1}{(2\pi)^3}
\int_{\Sigma \times \mathbf{CP}^3}
 H_3 \wedge \frac{1}{k} F^{(0)}_2
\wedge J \wedge J
\nn\\=&
T_{F1}\frac{1}{(2\pi)^3}
\frac{1}{k}
\int_{\Sigma \times \mathbf{CP}^3}
\left(
 d * \left(e^{-\Phi^{(0)}/2} F_6 \right)
-\frac{1}{2} d * \left(e^{-\Phi^{(0)/2}} \Phi F^{(0)}_6 \right)
+2\kappa_{10}^2 \mu_4 \delta^5 (x_{D4})
 \right) \wedge J \wedge J
\nn\\=&
T_{F1}\frac{1}{(2\pi)^3}
\frac{1}{k}
\int_{\Sigma \times \mathbf{CP}^3}
\left( (2\pi)^3 \frac{-1}{2\pi} J \delta^{AdS_4}(x_{D4})
+ d * \left(e^{-\Phi^{(0)}/2} F_6 \right)
 \right) \wedge J \wedge J
\nn\\=&
\frac{2\pi}{k}
+T_{F1}\frac{1}{(2\pi)^3}
\frac{1}{k}
\int_{\partial\Sigma \times \mathbf{CP}^3}
* \left(e^{-\Phi^{(0)}/2} F_6 \right)
\wedge J \wedge J \,. \label{anyonpairf1d4}
\end{align}
In the last line, the same argument goes as before, and the integral
for $*F_6$ can be dropped
when the distance between the F1 and the D4 source
are far enough.
Therefore, the anyonic phase of
the dressed baryons being made of D2-D4-D6 bound states is
\be\label{abphase2}
\Delta \phi_{D4D6}={2\pi \over k}(n'_6 N-n'_2 k) \; \mbox{mod} \; 2\pi
=2\pi n'_6 \lambda\;.
\ee
Note that only D6's fundamental strings contribute to the phase
nontrivially, and the AB phase of D4-D6 holographic anyons is
proportional to the 't Hooft coupling, not its inverse like the D0-D2
case.

For more generic case with fractional branes or fractional fluxes
wrapping on the cycles of $\mathbf{CP}^3$ such as the ones considered
in \cite{Aharony:2008gk}, we may obtain more varieties of anyonic phases.

In summary: From linearized supergravity analysis, we show that the
dressed baryons, which are either D0-D2 or D4-D6 bound states, are the candidates of the anyons for the dual Chern-Simons theory on the boundary.  Moreover, the fractional phases are proportional to either the 't Hooft coupling (for D4-D6) or its inverse (for D0-D2) so that they could persist even in the large $N$ and $k$
limit. This is in contrast to the anyonic pairs considered in the
previous section.

    Up to this point, two remarks are in order:
\begin{flushleft}
  \underline{D-brane solutions for the holographic anyons}
\end{flushleft}

 In the above, we have only shown that there are possible holographic
 anyon candidates as the spiky magnetic wrapped branes, however, we
 still need to solve these configurations from the field equations for
 the probe wrapped branes. Also, we also like to know if these
 holographic anyons are BPS objects or not. We will consider these
 issues in Section \ref{sec:BPS_cond}.

\begin{flushleft}
  \underline{Dressed baryons as anyons in $AdS_5 \times S^5$ ?}
\end{flushleft}

   One may wonder if we can construct the similar anyonic baryon dressed by the dissolved D-strings in $AdS_5 \times S^5$ case. The answer is no. This can be easily seen from the following relevant field equation
\be\label{ads5a}
d*(e^{\Phi^{(0)}} \tilde{F_3})-F_5^{(0)}\wedge H_3- \kappa^2_{10} \mu_5 \tilde{B}_2 \wedge \tilde{B}_2 \delta^4(x_{D5})=0
\ee
in which the source D5-brane is wrapping the $S^5$ with $N$ fundamental strings
stretching out to the $AdS_5$ boundary.
Note that the Poincar\'e dual 4-form $\delta^4(x_{D5})$ locates the D5-branes at a point inside  the spatial part of $AdS_5$, i.e., it is proportional to the spatial part of the volume form of $AdS_5$. From \eq{ads5a} we will obtain the induced anyonic phase for the probe fundamental string is
\be
\Delta \phi \sim \int_{\Sigma \times S^5} dA_1 \wedge dA_1 \delta^4(x_{D5})
\ee
However, the above integral is zero because $\delta^4(x_{D5})$ is
proportional to the spatial part of the volume form of $AdS_5$ and
$dA_1 \wedge dA_1$ is a 4-form inside the $S^5$ part, and then they are
trivial in the above integral.
Adding other branes wrapping the cycles on $S^5$ will not
 change the result. Therefore, there is no analog for anyonic QCD
 dressed baryon as in the ABJM theory.

\section{Dual operators for holographic anyons}

In 2+1 dimensions, the anyons can be realized by attaching the
magnetic fluxes to the electrically charged particles.
This can be simply realized via a Chern-Simons theory as a low energy
effective theory of strongly coupled Landau fermions, and its brane
construction has been considered in \cite{Fujita:2009kw} by using a
D4-brane wrapping on $\mathbf{CP}^1$ of $\mathbf{CP}^3$ to realize the edge states of
the Fractional Quantum Hall Effect (FQHE).
There, the anyon is identified as the fundamental string attached to the
edge.
However, this is not the anyon considered here. In our case, the
anyons are holographically realized as the magnetized particle-like
branes in the original $AdS_4 \times \mathbf{CP}^3$ background, and
shall be realized as the dressed baryons in the dual
Chern-Simons-Matter theory.

 We recall the discussions on the particle-like branes in the original
 ABJM paper \cite{Aharony:2008ug}, see also \cite{Park:2008bk}.
The key point is to identify the RR symmetry as the global baryon number
 symmetry in the dual Chern-Simons-Matter theory, namely, the symmetry
 currents are dual to each other,
\be
J_b \longleftrightarrow J=k n_0+N n_4
\ee
where $n_0$ and $n_4$ are the charges of D0 and wrapped D4-branes as
 defined before.
So, the $n_0$ D0-branes will be schematically dual to the chiral
 operators $C^{k n_0}$ where $C_I$ is bosonic bi-fundamental matter fields in the
 ${\bf 4}$ representation of the $SU(4)_R$, and the $n_4$ D4-branes will be
 dual to di-baryon operator $[\det(C)]^{n_4}$.
On the other hand, the wrapped D2 and D6-branes
need to be attached with $k$ or $N$ fundamental strings whose
 other ends are on the boundary, and they
 could thus be dual to the baryons on the field theory side.
  According to
 the supergravity analysis, one should dress the dual of the baryons
by the magnetic flux to have the nontrivial anyonic phase, and the
magnetic fluxes should be the dissolved D0 branes.
These dressed baryons are the bound state of baryonic spiky branes and
 particle-like branes in the supergravity side, and then
they should correspond to the bound states of the baryons and the
 chiral operators such as $C^{k n_0}$ and $[\det(C)]^{n_4}$.
Then, the question is how could we have
nontrivial AB phase when winding one dressed baryon around the other?

The key point to answer the above question lies in the fact that
the baryon number discussed above coincides with the anti-diagonal
$U(1)_{b}$ of $U(N) \times U(N)$, thus the chiral operators $C^{k
n_0}$ and $[\det(C)]^{n_4}$ carrying  nonzero baryon number are not
gauge invariant.
Instead,  one needs to make them gauge invariant by attaching the
appropriate 't Hooft disorder operator, which can also be defined by
the large gauge transformation generated by the center of the gauge
group along a given contour ${\cal C}$, known as the 't Hooft
loop  \cite{tHooft:1977hy,Moore:1989yh,Itzhaki:2002rc}.
More explicitly, the $C^{k}$ should be attached by a 't Hooft disorder
operator  in the  $(\text{Sym}( \mathbf{N}^k) , \text{Sym}(
\bar{\mathbf{N}}^k) )$ representation, denoted by $T_{0}$
\cite{Aharony:2008ug}, and  $\det(C)$ by the $U(1)$ 't Hooft operator
which is equivalent to  the Wilson line $e^{iN\int_x^{\infty} a_b}$
denoted by  $W_{4}$ \cite{Park:2008bk}\footnote{
We should comment on one subtlety here.
$W_4$ is equivalent to a monopole with a fractional $U(1)_{\tilde b}$
charge.
In order for this monopole monopole with a fractional charge
to be allowed, we need
to identify the diagonal $U(1)$ to be the center, and then the gauge
group is essentially to be $[U(N) \times U(N) ]/ U(1)_{\tilde b}$.
This change would cause a problem in identifying the moduli space of
ABJM theory.
However, in the path-integral we can only include monopoles that are
compatible with fields in the fundamental representations of each $U(N)$,
and it does not lead to any significant difference
from the original setup.
So we here simply say that we have $\det(C)$ operator with original
ABJM setup.
See \cite{Park:2008bk} for details.},
where $a_b$ is the gauge field
of $U(1)_b$.
Even though the definition of the 't Hooft disorder operator via the
action of 't Hooft loop seems non-local, it was shown that
the fields in the ABJM theory cannot detect it when
winding around.
This is because $T_0$ causes the large gauge  transformation $(e^{2\pi i/N},
e^{2\pi i /N})$ and $W_4$ does $(e^{2\pi i /k},e^{2\pi i/k})$, and
therefore for the bifundamental matters in the ABJM theory
these phases cancel out\footnote{%
To be more precise, there is still difficulty in
invisibility of $T_0$ and then locality of $C^k$
in non-Abelian theory.
To define a good local $C^k$ operator, we would need to employ the
state-operator correspondence of CFT \cite{monopole}, and the
exact definition of our anyon operator in this manner will be left to
a future work.}.
 On the other hand, the 't Hooft disorder operators may
detect each other while winding around, and could be anyons.

Indeed, Itzhaki \cite{Itzhaki:2002rc} has shown that in $U(N)_k$
Chern-Simons theory without charged matters the 't Hooft operators are
equivalent to the Wilson lines in  $\text{Sym} (\mathbf{N}^k)$
representation \footnote{The equivalence does not hold if there are
charged matters in the theory, like the ABJM theory.},
and moreover, they are anyons.
This is because exchanging two 't Hooft operators is related to an Wilson
line in $\text{Sym} (\mathbf{N}^k)$ representation under the large
gauge transformation generated by the 't Hooft loop, and such a large
gauge transformation will yield a fractional AB phase, i.e., $\pi
k/N$.
However, in the ABJM theory it is not clear what is the gravity dual
of the single 't Hooft operator, otherwise it could be the holographic
anyon.
Instead the chiral primary-'t Hooft disorder operator bound
states dual to wrapped D0 and D4-branes are gauge invariant
configurations, and therefore are also invariant under the large gauge
transformation generated by the 't Hooft loop. That is, these bound
states cannot be anyons.
This seemingly negative result in search for the dual of holographic
anyons, but, is consistent with the supergravity analysis
eq.\eq{abphase1} which says that the fractional phase is absent if
there is D2 nor D6 charge in the bound state.

All the above
suggests that one needs to turn on either D2 or D6 charges to have
holographic anyons, that is, we need to consider baryon operators or
the dressed ones.

Now we turn to consider the possible candidates for the field theory
dual of D2-brane wrapping on $\mathbf{CP}^1 \subset \mathbf{CP}^3$
or D6 brane on the whole $\mathbf{CP}^3$.
This is dual to a baryon vertex in the field theory side, which binds
either $k$ fundamental fields $Q$'s or $N$ anti-fundamental
fields \footnote{It is anti-fundamental for D6 string since the
orientations of the fundamental strings for D2 and D6 are opposite.}
$Q^{\dagger}$'s to make a bound state.
In the IIB brane construction of the ABJM theory,
we may think that the new fundamental field $Q$ can be realized by the
open string between the probe D3 or D7 branes and the background D3-branes which are
separated by the NS5-branes into two parts, corresponding to the first
and second $U(N)$, respectively.
As suggested in ABJM \cite{Aharony:2008ug}, one may also consider these
$Q$ fields as the ends of Wilson lines that are dual to the
fundamental strings in the bulk.
The end points indeed transform as the fundamental representation and
would not carry the charge of the global $U(1)_b$.
So we here assume that $Q$ are not charged under $U(1)_b$, which is also
consistent with the supergravity analysis.

\begin{figure}[htb]
\begin{center}
\includegraphics[width=10em,clip]{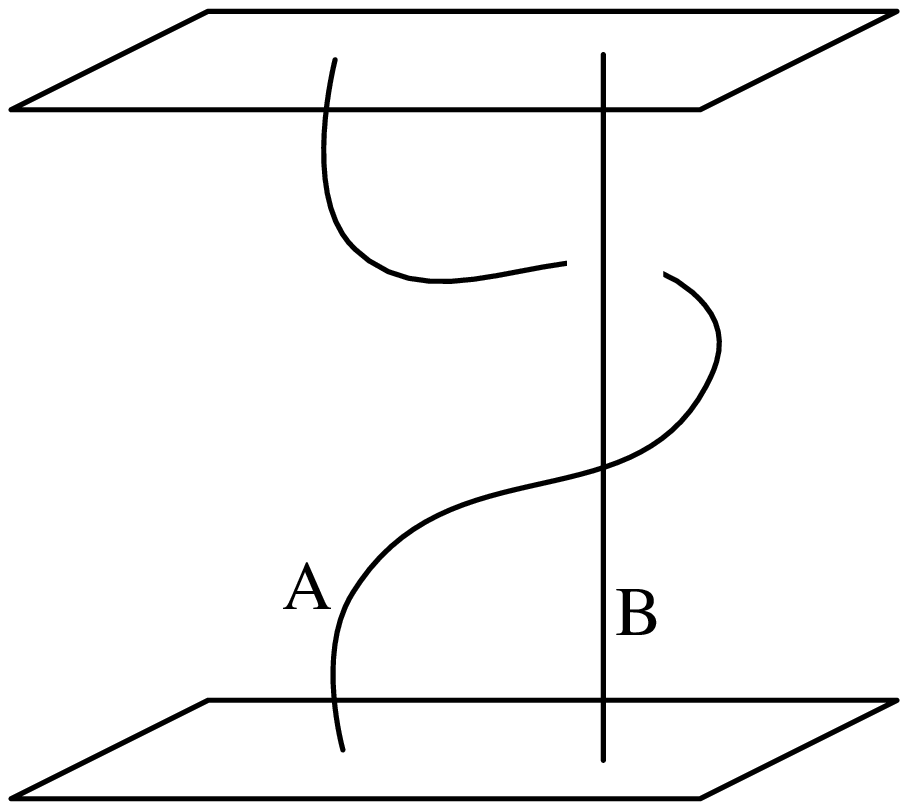}
\hspace*{6em}
\includegraphics[width=10em,clip]{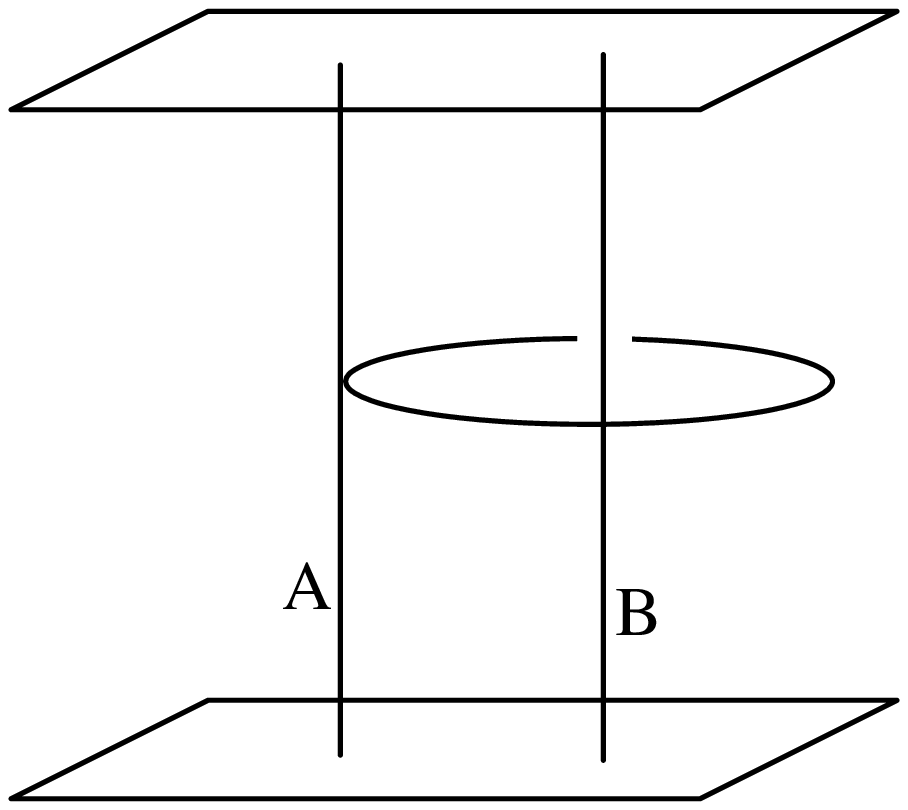}
\end{center}
\caption{Adiabatic exchange of two particles A and B (left). This is
  topologically equivalent to the right configuration with linking
  number 1 and A will get gauge-transformed if B is the 't Hooft
  operator.}
\label{anyons}
\end{figure}

Introducing D2-brane amounts to introducing baryonic bound state of
 $k$ $Q$-fields \footnote{%
Adopting the identification of $Q$ fields with D3-D3 (or $\bar Q$
 with D7-D3 string), one may specify the statistics of the ground
 states for the open string by counting the number of
 Neumann-Dirichlet boundary conditions, as in \cite{Witten:1998xy}.
It suggests that the ground state would be bosonic
 for 3-3 string and fermionic for 3-7 string.
We however do not pursue this issue in this paper.}, denoted by $Q^k$.
Naively, it seems that we need to introduce the attached 't Hooft
disorder operator as before to make such baryon gauge
invariant. However, this is not true because the dual D2-brane has
F-strings stretching to the boundary, and it is no longer just a
closed string state.
Since $Q$ is in the fundamental representation of the first $U(N)$
only, Chern-Simons action may provide magnetic fluxes attached to it,
and make it anyonic.
Or we may adopt the Wilson line interpretation of $Q$, and in this case
it also will have non trivial effect when two of them are exchanged.
However, due to the existence of the other charged matters, this
analysis is not easy to carry out.
So here we concentrate on the part of the anyonic phase we can
calculate unambiguously.
This treatment is also in line with the analysis in the supergravity
side, where only D2-D0 brane pair essentially contributes to the
anyonic phase to the leading order.
We then consider the baryon $Q^k$ dressed by the bound state of $C^k$ and
 $T_0$.
Since $Q^k$ is not invariant under the action of 't Hooft
loop,  when $Q^k$ goes around $T_0$, it gets the gauge
transformation (see Figure \ref{anyons})
\begin{align}
  Q^k \rightarrow  (e^{2i \pi\frac{1}{N}})^k Q^k=e^{\frac{2i \pi k}{N}} Q^k,
\end{align}
while if it goes around $W_4$, it will be
\begin{align}
  Q^k \rightarrow (e^{2i\pi\frac{1}{k}})^k
Q^k = Q^k \,.
\end{align}
So D2-brane baryon detects the existence of $T_0$ but not $W_4$. This
implies that $Q^k$--$C^k$--$T_0$ bound states are the holographic anyons
with the fractional phase equal to $n_2$ multiples of $2\pi k/N$, this is in
agreement with \eq{abphase1}.

  Similarly, D6-brane baryon is a bound state of $N$ $Q^{\dagger}$'s,
  denoted by $(Q^{\dagger })^N$, which would be a singlet of $SU(N)$.
  Now the situation is totally opposite to the D2-brane baryon:
  it acquires $e^{2\pi i N/k}$ phase factor from $W_4$ but the trivial one
  from $T_0$, i.e., it detects
  $W_4$ but not $T_0$.
We need to dress it by the $det(C)$--$W_4$
bound state to make anyon. Therefore, the $(Q^{\dagger
  })^N$--$det(C)$--$W_4$ bound states are the holographic anyons with
  the fractional phase equal to $n_6$ multiples of $2\pi N/k$ phase. Again,
  this fractional phase is captured by \eq{abphase2}.

  In summary: we assume the strings from the spiky D2 or D6 branes form the baryons, and by dressing them with the gauge invariant chiral primaries dual to the D0 or wrapped D4-branes, we can obtain the holographic anyons, with the fractional phases in agreement with the supergravity analysis.

Moreover, due to the non-trivial dressing, these holographic
anyons may have less supersymmetry than the chiral primaries, and
even are not BPS states\footnote{%
It has been pointed out that they would not be BPS configuration and
the reason is the following.
Since $Q^k$ corresponds to a flux of only one gauge group, $Q^k$ is
charged under only one gauge group, as stated in the text.
So it would sit in an angular momentum state and therefore there will
not be any BPS configuration of this kind of operator and usual chiral
operators.}.
Since it is not clear how to check the BPS condition for such a
composite operators, we will instead check it from their open string
duals in the next section.

\section{Constructing the holographic anyons}
\label{sec:BPS_cond}

In this section, we describe supersymmetric D0- and D2-brane
configurations in the ABJM background.
The anyonic pair in the supergravity side can be constructed by
these BPS configurations, though the bound states may break
supersymmetry.

  The Killing spinors in the ABJM background is summarized in
the Appendix \ref{appendix2}.
For our purpose in this section, it
is more convenient to work in the Poincar\' e coordinate for $AdS_4$.
Despite that, the Killing spinors given in \eq{killingsi}-\eq{killingsf}
are still too complicated to be used to solve the kappa symmetry
condition \eq{kappai}-\eq{kappaf} for the BPS embedding of the D-branes.
We have also tried to find BPS configurations of D4- and D6-branes,
but have not made it.
The setup and ansatz used there are also summarized in the Appendix C.

\subsection{D0-brane}
\label{sec:D0}

We start with considering D0-branes in the ABJM background
and find the BPS configurations.
In the ABJM paper, the chiral operators schematically represented as $C^{k}$
are identified with D0 brane in $AdS_4 \times \mathbf{CP}^3$
background.
These operators are in $\text{Sym}_{k}(\underline{\mathbf{4}})$
representation of $SU(4)_R$.
In the gravity side, this $SU(4)_R$ corresponds to the isometry of
$\mathbf{CP}^3$ and then BPS configuration would carry nontrivial
angular momenta in $\mathbf{CP}^3$.
We then consider D0 brane configurations rotating inside
$\mathbf{CP}^3$.
The Cartan subalgebras of $SU(4)_R$ correspond to the shifts in
$\chi, \vap_1$ and $\vap_2$ coordinates
and we thus turn on the angular momenta along these coordinates.

\paragraph{One-angular momentum case}

First we consider a general configuration,
where the D0 brane coordinates are
given by $x=y=0$,
$\theta_1, \theta_2, \alpha$ are all constant and
$\vap_1(t), \vap_2(t), \chi(t)$ and $r(t)$
 with the static gauge $\tau=t$.
Then the action is given by
\begin{align}
\label{D0_action_1ang_chi}
  S_{D0}=&
T_{D0} e^{-\Phi} \int dt \sqrt{-\det g}
-T_{D0} \int k \omega
\nn\\=&
T_{D0} \frac{k}{2} \int dt \left[\sqrt{H} - \frac{1}{2}
  \cos\alpha \dot\chi
-\cos^2\frac{\alpha}{2}\cos\theta_1 \dot\vap_1
-\sin^2\frac{\alpha}{2}\cos\theta_2 \dot\vap_2
 \right] \,,
\end{align}
where
\begin{align}
  H=& r^2 - \frac{\dot{r}^2}{r^2}
-A^2-B^2-C^2 \,,
\\
A=&
 \sin \frac{\alpha}{2} \cos \frac{\alpha}{2}
\left(\dot\chi+\cos\theta_1\dot\vap_1 - \cos\theta_2\dot\vap_2 \right) \,,
\\
B=&
 \cos \frac{\alpha}{2} \sin\theta_1 \dot\vap_1 \,,
\\
C=&
 \sin \frac{\alpha}{2} \sin\theta_2 \dot\vap_2
 \,.
\end{align}
Obviously, constant $\dot\chi, \dot\vap_1$ and $\dot\vap_2$
configuration solves the equation of motion and we will
assume this.

The $\kappa$ symmetry projector is given by
\begin{align}
  \Gamma=&
\frac{1}{\sqrt{H}} \Gamma_{11} \left(
r \Gamma_0
-\frac{\dot{r}}{r}\Gamma_3
+A\Gamma_5
+B \Gamma_7
+C \Gamma_9
\right) \,,
\end{align}
and the BPS condition is that
\begin{align}
  \Gamma \epsilon =&\epsilon \,,
\nn\\
\epsilon=& \mathcal{K} \mathcal{L} \mathcal{M} \epsilon_0 \,,
\end{align}
is solved by a constant spinor $\epsilon_0$.
We first take $r(t)=r_0$.
Inspired by the supersymmetry condition preserved by $M2$-branes
generating this background, we may impose a projection condition,
\begin{align}
  \hat\gamma \Gamma_3 \epsilon_0 =& \epsilon_0 \,,
\end{align}
the Killing spinor is a bit simplified by
$\mathcal{M} \epsilon_0 = \epsilon_0$.
After then we can take $r_0 \rightarrow 0$ and then now D0 brane is
sitting at the center of $AdS_4$.

Here we concentrate on the simplest case where only one of $A$, $B$ or
$C$ is non-zero.
All the cases go in parallel and then we consider $B=C=0$, that is,
$\dot\vap_1=\dot\vap_2=0$ case.
In this case the BPS condition is simplified
to be
\begin{align}
  i \Gamma_5 \Gamma_{11} \mathcal{K} \mathcal{L} \epsilon_0 =&
\mathcal{K} \mathcal{L} \epsilon_0 \,.
\end{align}
When $\theta_1=\theta_2=0$, $\Gamma_5 \Gamma_{11}$ commutes with
$\mathcal{K}$ and then we need to solve
\begin{align}
i  \Gamma_5 \Gamma_{11} \mathcal{L} \epsilon_0 =& \mathcal{L} \epsilon_0 \,.
\end{align}
By commuting $\Gamma_5 \Gamma_{11}$ with $\mathcal{L}$, we finally
arrive at
\begin{align}
  i \Gamma_5 \Gamma_{11} e^{-\frac{\chi}{4}(\hat\gamma\Gamma_{11}-\Gamma_{45})}
\epsilon_0 =& \epsilon_0 \,.
\end{align}
We then impose the further projection conditions
\begin{align}
  i \Gamma_5 \Gamma_{11} \epsilon_0 =& \epsilon_0 \,,
\qquad
\hat\gamma \Gamma_{11} \epsilon_0 = \Gamma_{45} \epsilon_0 \,,
\end{align}
to solve the BPS condition.
Let us count the number of the supersymmetry preserved by this
D0-brane.
Together with the previous projection condition and
(\ref{proj_cond_Killing}), we have imposed the conditions
\begin{gather}
  \hat\gamma \Gamma_3 \epsilon_0 = \epsilon_0 \,,
\qquad
\hat\gamma \Gamma_{11} \epsilon_0 = \Gamma_{45} \epsilon_0 \,,
\qquad
\Gamma_{67} \epsilon_0 = \Gamma_{89} \epsilon_0 \,,
\nn\\
  i \Gamma_5 \Gamma_{11} \epsilon_0 = \epsilon_0 \,.
\end{gather}
So this configuration is a $1/6$ BPS configuration.

\subsection{D2-branes}
\label{sec:spiky}

  Since the wrapped D2-branes carry RR 2-form charges which should be canceled by the
fundamental strings extending to the infinity. Similar story happened before for the wrapped D5-brane
as the dual baryons proposed in \cite{Witten:1998xy}. Soon it was realized that the whole configuration
can be realized as a spiky branes \cite{Imamura:1998gk,Callan:1998iq} in AdS space, quite similar to its flat space counterpart considered in \cite{Callan:1997kz,Gibbons:1997xz}.  Moreover, this configuration is BPS, in contrast to the non-BPS one considered in \cite{Brandhuber:1998xy} by simply attached the fundamental strings to the wrapped branes. Following the same reasoning, it suggests that D2-brane wrapping
on $\mathbf{CP}^1$ with $k$-strings attached and D6-brane wrapping
on the whole $\mathbf{CP}^3$ with $N$-strings attached will be BPS
when we replace the bunch of strings with a ``spike'' solutions on the
DBI action \cite{spike-solutions}. We now construct the BPS spiky
D2-branes here.
However, we also find that the spiky D2 with magnetic
flux satisfying \eq{d2B} does not solve the equations of motion. Thus,
we cannot have holographic anyon only from the spiky D2, instead we
need to use the bound states such as the ones of spiky D2 and D4 with
magnetic flux satisfying \eq{d4B}. Though this kind of holographic
anyons could be unstable.

\paragraph{Ansatz}

Suppose the D2-brane is wrapping on $\mathbf{CP}^1$ given by
 $\alpha=0$
 slice
of the $\mathbf{CP}^3$ and has a spike sourced by $k$-unit of the
electric charge.
The $\mathbf{CP}^1$ is parameterized as
\begin{equation}
  ds^2_{CP^1} =
\frac{1}{4} \left(
d\theta_1^2
+\sin^2 \theta_1 d\varphi_1^2
 \right) \,,
\end{equation}
and because of the factor $4$ in (\ref{eq:IIA-bg}),
now $AdS_4$ and $\mathbf{CP}^1$ has the same radius.
We take the static gauge
\begin{equation}
  \tau =t \,,
\qquad
 \sigma_1 =  \theta_1  \,,
\qquad
 \sigma_2 =  \varphi_1 \,,
\end{equation}
and the brane configuration is assumed to be given by $r(\theta_1)$.
We also turn on fluxes on the D2-brane, to be generic we have both electric and magnetic fluxes
\begin{align}
\mathcal{F} =&  2\pi \alpha' F
\nn\\ =&
E^1 d\tau \wedge d\sigma^1
+E^2 d\tau \wedge d\sigma^2
+B d\sigma^1 \wedge d\sigma^2
\,.
\end{align}

The DBI part of the action is
\begin{align}
  S_{DBI}=&
T_{D2} \int e^{-\Phi}
\sqrt{-\det (g+2\pi\alpha' F)}
\nn\\=&
T_{D2} \frac{k}{2}\tilde{R}^2 \int dt d\theta_1 d\vap_1
\, \sqrt{H} \,,
\\
H=& \sin^2 \theta_1 \left(r^2+r^{\prime 2} \right)
+r^2 \mathbf{B}^2
-\sin^2\theta_1 \mathbf{E}_1^2
-\left(1+ \frac{r^{\prime 2}}{r^2}\right) \mathbf{E}_2^2 \,,
\end{align}
where $\mathbf{B}= B / \tilde{R}^2$ etc.
This D2-brane also coupled to the background RR 2-form flux,
\begin{align}
  S_{WZ}=&
-T_{D2} \int \mathcal{F} \wedge k \omega
\nn\\=&
- T_{D2} \frac{k}{2} \tilde{R}^2
 \int \mathbf{E}_1 \cos\theta_1 dt d\theta_1 d\vap_1 \,.
\end{align}
Thus the equations of motions are
\begin{align}
    \label{eq:Asigma1-eom}
0=&    \partial_{t}
\left(
\frac{\sin^2\theta_1}{\sqrt{H}}\mathbf{E}_1
 \right)
-\partial_{\vap_1}
\left(
 \frac{r^2 \mathbf{B}}{\sqrt{H}}
\right) \,,
  \\
    \label{eq:Asigma2-eom}
0=&    \partial_{t}
\left(
\frac{1+\frac{r^{\prime 2}}{r^2}}{\sqrt{H}}\mathbf{E}_2
 \right)
+\partial_{\theta_1}
\left(
 \frac{r^2 \mathbf{B}}{\sqrt{H}}
\right)  \,,
\\
    \label{eq:Atau-eom}
0=& -    \partial_{\theta_1}
\left(
\frac{\sin^2\theta_1}{\sqrt{H}}\mathbf{E}_1
+\cos\theta_1 \right)
-\partial_{\vap_1}
\left(
 \frac{1+\frac{r^{\prime 2}}{r^2}}{\sqrt{H}}\mathbf{E}_2
\right)  \,,
\\
    \label{eq:r-eom}
0=&    \partial_{\theta_1}
\left(
\frac{r' \sin^2\theta_1 - \frac{r'}{r}\mathbf{E}_2^2}{\sqrt{H}}
 \right)
- \frac{r\sin^2\theta_1 + r \mathbf{B}^2 + \frac{r^{\prime 2}}{r^3} \mathbf{E}_2^2}{\sqrt{H}}
 \,.
\end{align}

\paragraph{$\kappa$-symmetry Projector}

For this D2-brane configuration,
$\kappa$-symmetry projector becomes
\begin{align}
  \Gamma=& \frac{1}{\sqrt{H}}
 \left[
r\sin\theta_1 \Gamma_{067}
-r' \sin\theta_1\Gamma_{037}
+\sin\theta_1 \mathbf{E}_1  \Gamma_7 \Gamma_{11}
-\mathbf{E}_2 \Gamma_6 \Gamma_{11}
+\frac{r'}{r} \mathbf{E}_2 \Gamma_3 \Gamma_{11}
-\mathbf{B} r\Gamma_0 \Gamma_{11}
\right]\,.
\end{align}

Now the BPS equation reads
\begin{align}
  \Gamma \mathcal{K} \mathcal{L} \epsilon_0 =&
\mathcal{K} \mathcal{L} \epsilon_0 \,,
\end{align}
where we have already assumed $\hat\gamma \Gamma_3 \epsilon_0 =
\epsilon_0$.
First consider
\begin{align}
  \mathcal{K}^{-1} \Gamma \mathcal{K}
=&
\frac{1}{\sqrt{H}} \left[
\left(
r \sin\theta_1 \Gamma_{067}
-r' \sin\theta_1 \Gamma_{037}
-\mathbf{E}_2 \Gamma_{6} \Gamma_{11}
+\frac{r'}{r}\mathbf{E}_2 \Gamma_3 \Gamma_{11}
+\mathbf{B}r \Gamma_0 \Gamma_{11}
\right)
e^{-\frac{\theta_1}{2}(\hat\gamma \Gamma_6 - \Gamma_7 \Gamma_{11})}
\right.\nn\\&\left.\hskip1cm
+\mathbf{E}_1 \sin\theta_1 \Gamma_7 \Gamma_{11}
\right] \,,
\end{align}
where we have used $\alpha=\theta_2=0$.

We here assume the following projection conditions on the constant
spinor $\epsilon_0$,
\begin{align}
  \hat\gamma \Gamma_3 \epsilon_+ =&  \Gamma^{012} \epsilon_0
  =\epsilon_0 \,,
 \\
 \Gamma_{067} \epsilon_0 =& u \epsilon_0 \,,
\\
 \Gamma_{03} \Gamma_{11} \epsilon_0 =& v \epsilon_0 \,,
\end{align}
where $u$ and $v$ are $\pm 1$.
The reason is the following.
 The first condition is
 the SUSY condition for the background $N$
M2-branes (or $N$ D2-branes) in the flat spacetime, and we have
 already imposed this condition for $t$-independence.
The next condition is a (local) BPS condition for D2-branes wrapping
on $\mathbf{CP}^1$ whose tangent space is given by $6,7$ directions.
The last one is the BPS condition for the fundamental string
stretching along the $r$-direction.
With this ansatz, it is easy to see that
\begin{align}
  \mathcal{L} \epsilon_0 =&
e^{\frac{\vap_1}{4}(v-u)\Gamma_0} \epsilon_0 \,.
\end{align}
When $\mathbf{B}=0$ or $u=v$, this factor is commuting with $\mathcal{K}^{-1} \Gamma \mathcal{K}$
and decouples from the BPS equation.
Next consider
$e^{-\frac{\theta_1}{2}(\hat\gamma \Gamma_6 - \Gamma_7 \Gamma_{11})}
\epsilon_0$
part.
Note that this factor commutes with $\mathcal{L}$ now.
Under the projection condition, it becomes
\begin{align}
  e^{-\frac{\theta_1}{2}(\hat\gamma \Gamma_6 - \Gamma_7 \Gamma_{11})}
\epsilon_0
=&
e^{-\frac{\theta_1}{2} (1-uv) \Gamma_{36}
} \epsilon_0 \,.
\end{align}
So we assume $u=v$ and then can take $\mathcal{L}=1$.
Further employing the projection conditions, the BPS equation becomes
\begin{align}
  \left[
ur \sin\theta_1 - \sqrt{H}
-(v r' - \mathbf{E}_1) \sin\theta_1 \Gamma_7 \Gamma_{11}
-\mathbf{E}_2 \Gamma_6 \Gamma_{11}
+\frac{r'}{r}\mathbf{E}_2 \Gamma_3 \Gamma_{11}
+\mathbf{B}r \Gamma_0 \Gamma_{11} \right] \epsilon_0 =0 \,.
\end{align}
Since $\Gamma_6 \Gamma_{11}, \Gamma_7 \Gamma_{11}$ and $\Gamma_0 \Gamma_{11}$ are not
commuting with the projection conditions, we need to set
all the coefficients to be zero, that is,
$\mathbf{E}_2=\mathbf{B}=0$ and
\begin{align}
  \mathbf{E}_1 = vr' \,.
\end{align}
The rest condition is
\begin{align}
  u r \sin\theta_1 = \sqrt{H} \,,
\end{align}
and this can be solved by $u=1$.
In order to fix the profile of the spike, we then consider the
equations of motion.
The first two equations (\ref{eq:Asigma1-eom}) and
(\ref{eq:Asigma2-eom}) are trivially satisfied.
The last two equations (\ref{eq:Atau-eom}) and (\ref{eq:r-eom})
lead the same equation
\begin{align}
  \partial_{\theta_1} \left(\frac{r'}{r}\sin\theta_1 \right)
=\sin\theta_1 \,,
\end{align}
which can easily be integrated and the solution is given by
\begin{align}
  r(\theta_1)=&
C_1 \frac{1}{\sin\theta_1} \left(\frac{1-\cos\theta_1}{\sin\theta_1}
\right)^{C_2} \,.
\end{align}
Here, one of the integration constant $C_2$
 will be set to one in the following by the plausible flux
distribution, and the other $C_1$ is the free moduli parameter for the
 radial position of the wrapped D2.
 Even though there
is no BPS solution for $\mathbf{B}\ne 0$ case, one may wonder if there
 is non-BPS solution for it.
However, it turns out that there is no
 spike solution of equations of motion for $\mathbf{E}_2=0$ but
 $\mathbf{B}\ne 0$ satisfying \eq{d2B}, i.e., $\mathbf{B}={m_2\over 2}
 \sin\theta_1$. To see this, one can first solve $\mathbf{E}_1$ from
 \eq{eq:Atau-eom}, and $\mathbf{B}$ is also given, then one can show
 that the remaining two equations \eq{eq:Asigma2-eom} and
 \eq{eq:r-eom} are not consistent with each other in solving the spike
 profile $r(\theta_1)$.

Note that by giving up on having a spike we can obtain a solution to
the equations of motion with a magnetic field.
To see this, first notice that we need to introduce $k$ charges
corresponding to the attached fundamental strings, as $k \delta^2 (x)$
term in the Wess-Zumino term.
Having this term allows us to set $A_0=\text{const.}$ consistently and
then we have $\mathbf{E}=0$.
Then together with $r=\text{const.}$, one finds that $\mathbf{B}
\propto \cos\theta_1$ solves the equations of motion.

\paragraph{Distribution of the electric flux}  In order to see Gauss
 law part of the equation of motion,
it is useful to rewrite the action using explicitly $A_\tau$.
We first consider the case with $\mathbf{E}_2=0$, $\mathbf{B}=0$
and $\mathbf{E}_1 = - \tilde{R}^{-2} 2\pi\alpha' \partial_{\theta_1}
A_t$.
Then the $A_t$ equation of motions is
\begin{align}
  \partial_{\theta_1} \Pi_{\theta_1} =&
 T_{D2} \frac{k}{2} 2\pi\alpha'  \sin\theta_1,,
\\
 \Pi_{\theta_1} =&  - T_{D2} \frac{k}{2} 2\pi\alpha'
\frac{\sin\theta_1 \mathbf{E}_1}{\sqrt{r^2+r^{\prime 2} -
 \mathbf{E}_1^2}}
\,,
\end{align}
where
\begin{align}
  \Pi_i =& \frac{\partial \mathcal{L}}{\partial (\partial_i A_t)}
\end{align}
is the conjugate momentum.
This corresponds to Gauss law $\nabla \cdot \mathbf{E} = \rho$ part of
the Maxwell equations and the integration of $\Pi$ defines a
conserved quantity and then
integration of the right hand side over the spatial volume (now
$\mathbf{CP}^1$) should give the total charge.

We here take the BPS spike solution,
\begin{align}
  r(\theta_1)=&
C_1 \frac{1}{\sin\theta_1} \left(\frac{1-\cos\theta_1}{\sin\theta_1}
\right)^{C_2} \,,
\nn\\
  \mathbf{E}_1 =& r'\,,
\quad \rightarrow \quad A_t(\theta_1) = r \,,
\end{align}
and we obtain
\begin{align}
   \Pi_{\theta_1}=&
- \frac{k}{4\pi} (C_2 - \cos\theta_1)
\,.
\end{align}
First we see the profile of the solution.
For $\theta_1$ goes to $\pi$, both of $r(\theta_1)$ and $A_t$
get divergent and then we conclude that the point charge is located at
$\theta_1 = \pi$.
Next for small $\theta_1$, both of $r(\theta_1)$ and $A_t$ behave as
$\theta_1^{C_2-1}$.
Therefore for the solution to be smooth on the other side of the
spike, $C_2 \geq 1$.
Finally, by integrating $\Pi$ over $\mathbf{CP}^1$ we find
\begin{align}
  \int \sin\theta_1 d\theta_1 d\vap_1 \Pi_{\theta_1}
=& - C_2 k \,.
\end{align}
This charge has to cancel the $k$ units of charge induced by the background,
and therefore we get $C_2=1$.
Thus the correct BPS solution with a plausible profile is given by
\begin{align}
r(\theta_1) =& A_t(\theta_1) =
 \frac{2 r_0}{1+\cos\theta_1}
 \,,
\end{align}
where $r_0 = C_1 / 2$ denote the position of the end of the spike at
$\theta=0$, i.e., the radial position of the wrapped D2-brane.

We then conclude that each D0-brane having an angular momentum
and D2-brane with a spike is a BPS configuration.
It however turned out that, within our ansatz,
the preserved supersymmetry by D0 and D2 branes are not compatible.
Furthermore there does not exist
BPS spike D2-branes with magnetic fluxes.
These facts imply that our dressed baryons, D0-D2 bound states,
are not BPS.

\section{Conclusion and Discussions}

In this paper, we have constructed the holographic anyons in the ABJM
theory  from the gravity, CFT and open string sides via AdS/CFT
correspondence.   The construction is more subtle than naively expected
in all three aspects because it is the nontrivial
generalization of the usual anyon constructed in the  $U(1)$
Chern-Simons effective theory.
In $U(1)$ case we attach the magnetic
flux to the
electron to make it anyon via the Chern-Simons coupling.
Similarly, here
we attach the nonabelian 't Hooft operator to the
baryon to make it anyonic.

We find two types of holographic anyons as the dressed baryons.
One is the D0-D2 bound states, and the other is the D4-D6 ones. The
anyonic phases
from gravity and CFT sides agree. For D4-D6, the anyonic phase is
proportional to
the 't Hooft coupling, and for D0-D2 its inverse.
Interestingly, these two pairs are not related by the usual Hodge
duality in ten dimensions, since it relates $C_1$ to $C_7$ and $C_3$
to $C_5$ but in the relation above the roles of D0 brane and D4 brane
are exchanged.
It has been suggested that this relation can be understood as a kind
of geometric duality inside $\mathbf{CP}^3$\cite{Park:2008bk}.
Moreover, by combing with the level rank duality we can transform one
anyonic phase to the other one, i.e.,
\be
N \longleftrightarrow k\;, \qquad n_0 \longleftrightarrow n_4\;, \qquad n_2\longleftrightarrow n_6\;,
\ee
and the anyonic phases are then switched as
\be
n_2 {2\pi k\over N} \longleftrightarrow  n_6{2 \pi N\over k}.
\ee
In the above, $n_p$ is the number of wrapped Dp-brane baryons in the
anyon bound states.
It is interesting to see if the combination of D0-D4 duality and
level-rank duality is related to the particle-vortex duality in the
quantum Hall system \cite{burgess}.
If this is the case, then D0-D2 and D4-D6 can be understood as the
particle-vortex dual pair of CFT's collective modes.

We also like to comment more on the agreement of the anyonic phases from
gravity and CFT sides since it suggests that the anyonic phases do not
run with the
coupling constant. This seemingly topological feature should be due to
the neglect of
the interactions between the BPS Dp-branes if they are far enough from
each other.
Especially, in the supergravity side, if the distance between two branes (or
strings and a brane) is not far enough, we may not be able to neglect
the dynamical part of the phase, and as the separation distance goes
to zero, the
phase will disappear. This behavior might correspond to the fact in the field theory side that the 't
Hooft loops, or Wilson loops, become unstable once we introduce the
fields which are not invariant under the center of the gauge group
\cite{tHooft:1977hy}. Therefore, the holographic anyonic phase is a
long-range property of these pairs.

As a by product, we also examine the Killing spinor equation for the
embedding branes wrapped over $\mathbf{CP}^3$. Though  we have found
the nontrivial BPS spiky wrapped D2-brane configuration, surprisingly
some expected BPS solution for the chiral primary such as wrapped D4
brane and spiky D6 brane are not found by the simple ansatz based on
symmetry argument.
Despite that, we have put down the details of our unsuccessful trials
and hopefully this will help for the further studies.
We also note that though
each of D0 brane and D2-brane with a spike is BPS,
their bound state is not due to unmatching of the supersymmetries they
preserve.
It thus suggest that our dressed baryons are not protected from quantum
corrections and it would appear very differently in either weak and
strong coupling regimes.
As noticed above, though, the anyonic phase (more precisely AB phase)
between D0 and D2 are stable when the distance between them are far
enough.
It is then also interesting to investigate whether there exist BPS dressed
baryon states in ABJM background.

We hope our results will inspire more studies on the connection
between string theory and other branches of physics via AdS/CFT
correspondence. It is also interesting to see if one can find the
holographic anyons in the other holographic duals, and moreover,
consider the dynamical consequences of these anyons, such as the
implementation of topological quantum computing.

\bigskip
\bigskip
\bigskip
\section*{Acknowledgments}
  This work is supported by Taiwan's NSC grant 097-2811-M-003-012 and 97-2112-M-003-003-MY3.
We also thank the support of NCTS.

\appendix

\section{Massive fluctuations}
\label{appendix1}

We will show  that $F_2$ and $H_3$ are massive fields in the AdS bulk. This can seen most easily from the relevant field equations:
\bea
& & d*F_2 - H_3 +4\pi \delta^3(x)=0\;,\\
& & d*H_3+F_2=0\;.
\eea
From the above, we will obtain
\be
d*d*H_3+H_3=4\pi \delta^3(x),
\ee
or
\be
\Delta H_3-H_3=-4\pi \delta^3(x).
\ee
Obviously, it is a massive field, so is $F_2$.

\section{Killing spinors and supersymmetric embeddings}
\label{appendix2}
The ABJM geometry in the string frame metric ($\ell_s=1$) is
\begin{align}
  \label{eq:IIA-bg}
  ds^2=& \tilde{R}^2 \left(
ds^2_{AdS_4} + 4 ds^2_{\mathbf{CP}^3}
\right) \,,
\\
  ds^2_{AdS_4} =&
r^2 \left(-dt^2 + dx^2 + dy^2 \right)
+ \frac{dr^2}{r^2} \,,
\\
  \label{CP3-coord}
ds^2_{\mathbf{CP}^3}=&
\frac{1}{4} \left[
d\alpha^2
+\sin^2 \frac{\alpha}{2}  \cos^2 \frac{\alpha}{2}
   \left(d\chi + \cos \theta_1 d\varphi_1
               - \cos \theta_2 d\varphi_2  \right)^2
\right.\nn\\&\left.\hskip1cm
+\cos^2 \frac{\alpha}{2}
   \left(d\theta_1^2 + \sin^2 \theta_1 d\varphi_1^2 \right)
\right.\nn\\&\left.\hskip1cm
+\sin^2 \frac{\alpha}{2}
   \left(d\theta_2^2 + \sin^2 \theta_2 d\varphi_2^2 \right)
\right]
\\
(0 \leq & \alpha, \theta_1, \theta_2 \leq \pi \,,
\qquad 0 \leq \vap_1, \vap_2 < 2\pi \,,
\qquad 0 \leq \chi < 4\pi )
\\
\tilde{R}^2 =& \frac{R^3}{4k} \,,
\qquad
\frac{R^3}{k} = 2^{5/2} \pi \sqrt{N\over k} \,,
\\
F_4 =& \frac{3}{8}R^3 d\Omega_{AdS_4} \,,
\quad
F_2 = k d\omega \,,
\quad
e^{2\Phi} = \frac{R^3}{k^3} \,,
\\
  \omega =&
\frac{1}{4}
\left(\cos\alpha d\chi
+2\cos^2\frac{\alpha}{2} \cos\theta_1 d\varphi_1
+2\sin^2\frac{\alpha}{2} \cos\theta_2 d\varphi_2 \right)
\,.
\end{align}
In particular, we have chosen the Poincare coordinate for the $AdS_4$, which is more
convenient for the Killing spinor analysis.

This background has the following vielbein:
\begin{align}
&  e^0 = \tilde{R} r dt \,,
\qquad
  e^1 = \tilde{R} r dx \,,
\qquad
  e^2 = \tilde{R} r dy \,,
\qquad
  e^3 =- \frac{\tilde{R}}{r} dr \,,
\nn\\ &
e^4 = \tilde{R} d\alpha \,,
\qquad
e^5 = \tilde{R} \sin \frac{\alpha}{2}  \cos \frac{\alpha}{2}
   \left(d\chi + \cos \theta_1 d\varphi_1
               - \cos \theta_2 d\varphi_2  \right)
\nn\\ &
e^6 =
\tilde{R}\cos \frac{\alpha}{2} d\theta_1
\,, \qquad
e^7 = \tilde{R} \cos \frac{\alpha}{2} \sin \theta_1 d\varphi_1
\nn\\ &
e^8 =
\tilde{R} \sin \frac{\alpha}{2} d\theta_2
\,, \qquad
e^9 = \tilde{R} \sin \frac{\alpha}{2} \sin \theta_2 d\varphi_2
\,.
\end{align}

\paragraph{Killing spinor}

This background turns out to have the following
Killing spinor
\begin{align}\label{killingsi}
  \epsilon =& \mathcal{K} \mathcal{L} \mathcal{M} \epsilon_0 \,,
  \\
  \mathcal{K} =&
e^{-\frac{\alpha}{4}(\hat\gamma \Gamma_4 - \Gamma_5\Gamma_{11})}
e^{-\frac{\theta_1}{4}(\hat\gamma \Gamma_6 - \Gamma_7\Gamma_{11})
+\frac{\theta_2}{4}(\Gamma_4 \Gamma_8 + \Gamma_5\Gamma_{9})} \,,
\\
\mathcal{L} =&
e^{\xi_1 \hat\gamma \Gamma_{11}+\xi_2 \Gamma_{45}+\xi_3
  \Gamma_{67}+\xi_4 \Gamma_{89} } \,,
\\
\mathcal{M} =&
r^{1/2}
\left[
\frac{1+\hat\gamma\Gamma_3}{2}
+\left(
  t\Gamma_0 + x\Gamma_1 + y \Gamma_2 + \frac{1}{r}\Gamma_3 \right)
\Gamma_3 \frac{1-\hat\gamma\Gamma_3}{2}
\right] \,, \label{killingsf}
\end{align}
where
\begin{equation}
  \xi_1=-\frac{\chi}{8} - \frac{\vap_1}{4} \,,
\quad
  \xi_2=\frac{\chi}{8} - \frac{\vap_2}{4} \,,
\quad
  \xi_3=-\frac{\chi}{8} + \frac{\vap_1}{4} \,,
\quad
  \xi_4=\frac{\chi}{8} + \frac{\vap_2}{4} \,.
\end{equation}
Note that
\be
{\cal M}\epsilon_0=\epsilon_0
\ee
if $\hat\gamma \Gamma_3 \epsilon_0 = \epsilon_0$.

And the dilatino condition gives
projection condition for the constant spinor
\begin{equation}
\label{proj_cond_Killing}
  (\hat\gamma \Gamma_{11} + \Gamma^{45}+\Gamma^{67}+\Gamma^{89})
  \epsilon_0=0
\,.
\end{equation}
  Moreover, these four Gamma matrices commute with each other and have their squares to be $-1$ so we can choose
\be
\hat{\gamma} \Gamma_{11}\epsilon_0=is_0\epsilon_0, \qquad \Gamma_{45}\epsilon_0=is_1\epsilon_0,\qquad
\Gamma_{67}\epsilon_0=is_2\epsilon_0, \qquad \Gamma_{89}\epsilon_0=is_3\epsilon_0,
\ee
where $s_i\in{\pm 1}$ and satisfy $\sum_i s_i=0$ so that the background is found to preserve $\mathcal{N}=6$ supersymmetry.
We can choose another matrix commuting with all of the above as
$\hat\gamma \Gamma_3=\Gamma^{012}$.
We may write the eigenvalue of this as $i s_5$ and then the 32
component spinor is specified by the set of the eigenvalues
$(s_0, s_1, s_2, s_3, s_4, s_5)$.

\paragraph{The projector}

The $\kappa$-symmetry projector for a Dp-brane with world-volume gauge field strength $\mathcal{F}_2$  in a Lorentzian background is given in
\cite{Bergshoeff:1996tu,Bergshoeff:1997kr} as
 \begin{align}\label{kappai}
   \Gamma=&
\frac{\sqrt{|g|}}{\sqrt{|g+\mathcal{F}|}}
\sum_{n=0}^\infty
\frac{1}{2^n n!} \gamma^{i_1 j_1 \cdots i_n j_n}
 \mathcal{F}_{i_1 j_1} \cdots \mathcal{F}_{i_n j_n} J_{(p)}^{(n)}
\,,
\\
J_{(p)}^{(n)}=& (\Gamma_{11})^{n+\frac{p-2}{2}} \Gamma_{(0)} \,,
\\
\Gamma_{(0)}=& \frac{1}{(p+1)!} \frac{1}{\sqrt{|g|}}
\varepsilon^{i_1 \cdots i_{p+1}} \gamma_{i_1 \cdots i_{p+1}}\,, \label{kappaf}
 \end{align}
where $i_1, \cdots$ are the world-volume indices,
$\gamma_i = \partial_i X^\mu \gamma_\mu$ is the pull-back of the
curved-space gamma matrices $\gamma_\mu$ onto the world-volume,
and $\gamma_\mu = e^{a}_{\mu}\Gamma_a$.

By using this projector, the BPS condition for the embedding is given
by
\begin{align}
  \Gamma \epsilon = \epsilon \,.
\end{align}

\section{Some trial for BPS D4 and spiky D6 brane configurations}
\label{sec:trial}

Apart from the D0 and D2-brane cases, we have also tried to solve the
BPS conditions for D4 and D6-brane cases.
Though we have not found BPS configurations, we here note our setup
and ansatz for future reference.

\subsubsection{D4-branes}
\label{sec:D4}

We consider a D4-brane wrapping on $\mathbf{CP}^2$.
In the original background given by $\bR^{1,2} \times \bC^4 /
\bZ_k$,
the would-be-$\mathbf{CP}^2$ can be regarded as a flat three plane
through the origin of $\bC^4$.
We then take $z_4=0$, which leads $\mathbf{CP}^2:\,
\theta_2=0 \,, \quad \vap_2=0$.

The configuration we consider is
\begin{center}
  \begin{tabular}{r|cccccccccc}
 & $t$ & $x$ & $y$ & $r$ & $\alpha$ & $\chi$
& $\theta_1$ & $\vap_1$ & $\theta_2$ & $\vap_2$ \\ \hline
D4 & $\circ$ &   &  & $r$ & $\circ$ & $\circ$
 & $\circ$ & $\circ$ & $0$ & $0$
  \end{tabular}
\end{center}
with $F_{\alpha\chi}$ and $F_{\theta_1 \vap_1}$ turned on.
Thus the DBI action is
\begin{align}
  S_{D4, DBI} =
T_{D_4} \tilde{R}^5 \frac{k}{2\tilde{R}}
\int dt d\alpha d\chi d\theta_1 d\vap_1 \, r\,
\cos^3 \frac{\alpha}{2} \sin \frac{\alpha}{2} \sin\theta_1
\sqrt{H} \,,
\end{align}
where
\begin{align}
\label{d4H}
H:=& (1+B_1^2)(1+B_2^2)\;,
\qquad  B_1 = \frac{2\pi\alpha' \, F_{\alpha\chi}}{\tilde{R}^2 \cos \frac{\alpha}{2} \sin \frac{\alpha}{2} } \,,
\qquad
B_2 =&
\frac{2\pi\alpha'\, F_{\theta_1\vap_1}}{\tilde{R}^2 \cos^2  \frac{\alpha}{2} \sin \theta_1 } \,.
\end{align}
Since we have not turned on any electric fields on D4-brane, the
Wess-Zumino term does not exist.

 The equations of motion of $B_1$ and $B_2$
are reduced to
\begin{align}
\partial_\alpha
\left(
\cos^2 \frac{\alpha}{2}
\frac{B_1 (1+B_2^2)}{\sqrt{H}}
\right) = 0 \,,
\qquad
\partial_{\theta_1}
\left(
 \frac{B_2 (1+B_1^2)}{\sqrt{H}}
\right) = 0 \,.
\end{align}
These equations are solved by
\be
B_1^2={C_1^2(C_2^2-1)\over C_1^2 -\cos^4{\alpha\over 2}}\;, \qquad B_2^2={C_2^2\over C_2^2-1} ({C_1^2 \over \cos^4{\alpha\over 2}}-1)
\ee
where $C_1(\theta_1)$ is an arbitrary function of $\theta_1$ only, and $C_2(\alpha)$ depends only on $\alpha$, instead.

\paragraph{BPS conditions}

The $\Gamma$ projector is given by
\begin{align}
  \Gamma=
\frac{1}{\sqrt{H}}
\left(
(\Gamma_{11}\Gamma_{45}-B_1)\Gamma_{067}(1-\sin{\alpha\over 2} \cot\theta_1 \Gamma_{57})-B_2 \Gamma_{045} +B_1 B_2 \Gamma_{11} \Gamma_0
\right)
\end{align}
and the kappa symmetry condition
\be
\label{BPS_eq_D4_1}
\Gamma {\cal K L}\epsilon_0={\cal K L}\epsilon_0
\ee
where we have chosen the projection condition
$\hat{\gamma}\Gamma_3\epsilon_0=\epsilon_0$ such that ${\cal
  M}\epsilon_0=\epsilon_0$.

We then assume the following projection condition
\begin{align}
  \hat\gamma \Gamma_{11} \epsilon_0
=& \beta_1 \Gamma_{45} \epsilon_0
= \beta_2 \Gamma_{67} \epsilon_0
= \beta_3 \Gamma_{89} \epsilon_0
\end{align}
and $\beta_i=\pm 1$ and the dilatino condition
(\ref{proj_cond_Killing}) requires that $\beta_1 + \beta_2 + \beta_3 =-1$.
Then we find that
\begin{align}
  \mathcal{L} \epsilon_0 =&
e^{(\beta_3 \xi_1 + \beta_1 \beta_3 \xi_2 + \beta_2 \beta_3 \xi_3 +
  \xi_4)\Gamma_{89}} \epsilon_0 \,,
\end{align}
and since $\Gamma_{89}$ commutes with both of $\mathcal{M}$ and
$\Gamma$, in (\ref{BPS_eq_D4_1})
 the $\mathcal{L}$ part trivially cancel on the both hands sides.

  We then need to commute $\Gamma$ with ${\cal K}$ to solve the BPS
 condition.
After some algebra, we  arrive at
\begin{align}
   \label{eq:KGK_D4}
& \sqrt{H} \mathcal{K}^{-1} \Gamma \mathcal{K}
\nn\\=&
- \Gamma_0 \Gamma_{11} \Gamma_{4567}
 \left(
 1-\sin\frac{\alpha}{2} \cot\theta_1
  \left(\cos\frac{\alpha}{2} \Gamma_{57} e^{\frac{\theta_1}{2}\Gamma_7
    \Gamma_{11}} - \sin\frac{\alpha}{2} \Gamma_7 \Gamma_{11} \right)
 \right)
\nn\\&
-B_1 \Gamma_{067}
 \left( 1+ \sin^2 \frac{\alpha}{2} \cot \theta_1 \Gamma_7 \Gamma_{11}
 \right)
 e^{- \frac{\theta_1}{2}( \hat\gamma \Gamma_6 - \Gamma_7 \Gamma_{11})}
\nn\\&
+B_1 \Gamma_{056} \sin\frac{\alpha}{2} \cos\frac{\alpha}{2} \cot\theta_1
e^{- \frac{\theta_1}{2} \hat\gamma \Gamma_6}
\nn\\&
-B_2 \Gamma_{045}
 \left(\cos\frac{\alpha}{2} - \sin\frac{\alpha}{2} \hat\gamma \Gamma_4
 e^{- \frac{\theta_1}{2} \hat\gamma \Gamma_6} \right)
 \left(\cos\frac{\alpha}{2} + \sin\frac{\alpha}{2} \Gamma_5 \Gamma_{11}
 e^{ \frac{\theta_1}{2} \Gamma_7 \Gamma_{11} } \right)
\nn\\&
-B_1 B_2 \Gamma_0 \Gamma_{11}
 \left(\cos\frac{\alpha}{2} - \sin\frac{\alpha}{2} \hat\gamma \Gamma_4
 e^{\frac{\theta_1}{2} \hat\gamma \Gamma_6} \right)
 \left(\cos\frac{\alpha}{2} + \sin\frac{\alpha}{2} \Gamma_5 \Gamma_{11}
 e^{- \frac{\theta_1}{2} \Gamma_7 \Gamma_{11} } \right)
 e^{- \frac{\theta_1}{2}( \hat\gamma \Gamma_6 - \Gamma_7 \Gamma_{11})}
\,.
\end{align}

Let us first consider $(\beta_1,\beta_2,\beta_3)=(-,+,-)$ case.
For this choice, the BPS equation becomes
\begin{align}
\epsilon_0 =&
\frac{1}{\sqrt{H}}\left[
-(1+ B_1B_2 \cos\alpha) \Gamma_{0}\Gamma_{11}
-B_1 \sin^2\frac{\alpha}{2} \cot\theta_1 \Gamma_{0457} \Gamma_{11}
\right.\nn\\&
+\left(\cos\frac{\alpha}{2}\sin\frac{\alpha}{2} \cot\theta_1
\cos\frac{\theta_1}{2}
- B_1 B_2 \sin\alpha \sin\frac{\theta_1}{2} \right)\Gamma_{057}
\Gamma_{11}
\nn\\&
-\left(B_1 \cos\frac{\alpha}{2}\sin\frac{\alpha}{2} \cot\theta_1
\sin\frac{\theta_1}{2} + B_2 \sin\alpha \cos\frac{\theta_1}{2}
\right)\Gamma_{04} \Gamma_{11}
\nn\\&
+\left(\cos\frac{\alpha}{2}\sin\frac{\alpha}{2} \cot\theta_1
\sin\frac{\theta_1}{2} + B_1 B_2 \sin\alpha \cos\frac{\theta_1}{2}
\right) \Gamma_{05}
\nn\\&
-\sin^2\frac{\alpha}{2} \cot\theta_1 \Gamma_{07}
+(B_1 - B_2\cos\alpha) \Gamma_{045}
\nn\\& \left.
+ \left(B_1 \cos\frac{\alpha}{2}\sin\frac{\alpha}{2} \cot\theta_1
\cos\frac{\theta_1}{2} - B_2 \sin\alpha \sin\frac{\theta_1}{2}
\right)\Gamma_{056}
\right] \epsilon_0
\end{align}
We have $\Gamma_{07}$ term whose coefficient does not include $B$'s.
For BPS solutions to exist, this $\Gamma_{07}$ should be projected to
be either $\mathbf{1}$ or one of the other gamma matrices.
However, any of this choice will not be compatible with the projection
conditions we have already imposed and then will break all the
supersymmetry.
We thus see that there is no BPS solution.
We have also checked the other two cases,
$(\beta_1,\beta_2,\beta_3)=(+,-,-)$ and
$(\beta_1,\beta_2,\beta_3)=(-,-,+)$, and have arrived at the same
structure and not found any BPS configuration based on this ansatz.

\subsubsection{$\mathbf{CP}^1 \times \mathbf{CP}^1$ embedding}

By setting $\theta_1=\theta_2:=\theta$ and $\vap_1=\vap_2:=\vap$, the $\mathbf{CP}^3$ is reduced to
$\mathbf{CP}^1 \times \mathbf{CP}^1$ of the
equal radius, we then wrap D4-brane on it.

We may turn on two independent magnetic fields $F_{\alpha\tilde\chi}$
and $F_{\theta \vap}$ and then the action is (Wess-Zumino part
again vanishes)
\begin{align}
  S_{D4, DBI} =& T_{D_4} \int e^{-\Phi} \sqrt{-\det (g+\mathcal{F})}
\nn\\=&
T_{D_4} \tilde{R}^5 \frac{k}{2\tilde{R}} \int
d\alpha d\tilde\chi d\theta d\vap \, r\,
\sin \alpha \sin\theta \sqrt{H} \,,
\\
& B_1 = \frac{2\pi \alpha' F_{\alpha\tilde\chi}}{\tilde{R}^2 \sin\alpha}
\,, \qquad
B_2 = \frac{2\pi \alpha' F_{\theta \vap}}{\tilde{R}^2 \sin\theta}
\,,
\end{align}
where $\tilde{\chi}=\chi/2$ and $H$ is the same as \eq{d4H}.

The equations of motion are (by choosing the gauge $F_{\alpha\tilde\chi}=\partial_{\alpha}A_{\tilde\chi}$ and $F_{\theta \vap}=\partial_{\theta}A_{\vap}$)
\begin{align}
    \partial_\alpha \left(\frac{B_1(1+B_2^2)}{\sqrt{H}}\right) =0
\,, \qquad
    \partial_{\theta_1} \left(\frac{B_2(1+B_1^2)}{\sqrt{H}}
\right) =0 \,.
  \end{align}
 These are solved by
 \be
 B_1^2={C_1^2 \over 1-C_1^2}\;, \qquad B_2^2={C_2^2\over 1-C_2^2}\;,
 \ee
 where $C_1=C_1(\theta)$, $C_2=C_2(\alpha)$ are arbitrary functions.

  The $\Gamma$ projector is then given by
 \be
 \Gamma={1\over \sqrt{H}}(\Gamma_{11}\Gamma_{045\tilde{6}\tilde{7}}-B_1\Gamma_{0\tilde{6}\tilde{7}}-B_2\Gamma_{045}+B_1B_2 \Gamma_{11}\Gamma_0)\;,
 \ee
where
\be
\Gamma_{\tilde{6}}:=\cos{\alpha\over 2} \Gamma_6 + \sin{\alpha\over 2} \Gamma_8\;, \qquad \Gamma_{\tilde{7}}:=\cos{\alpha{}\over 2} \Gamma_7 + \sin{\alpha{}\over 2} \Gamma_9\;.
\ee

We will impose $\hat\gamma \Gamma_3 \epsilon_0 = \epsilon_0$ condition
as before.
We need to impose further conditions for simplicity.
The simplest projection condition here is to choose
\begin{align}
  \hat\gamma \Gamma_{11} \epsilon_0 =&
-\Gamma_{45} \epsilon_0 =
-\Gamma_{67} \epsilon_0 =
\Gamma_{89} \epsilon_0 \,,
\end{align}
which leads $\mathcal{L} \epsilon_0 = \epsilon_0$.
With this choice,  $\mathcal{K} = e^{-\frac{\alpha}{4}(\hat\gamma
  \Gamma_4-\Gamma_5 \Gamma_{11})} e^{-\frac{\theta}{2}(\hat\gamma
  \Gamma_6-\Gamma_{48})}$ on $\epsilon_0$, and then
BPS equation is reduced to
\begin{align}
  \epsilon_0 =&
\frac{\Gamma_0}{\sqrt{H}}
\left[
\left\{
 e^{-\theta \Gamma_{48}}
  \left(\cos^2\frac{\alpha}{2}+ \sin^2\frac{\alpha}{2} B_1 B_2 \right)
-e^{\theta \hat\gamma \Gamma_6}
  \left(\sin^2\frac{\alpha}{2}+ \cos^2\frac{\alpha}{2} B_1 B_2 \right)
\right\} \Gamma_{11}
\right.\nn\\&\hskip2cm
+\left\{
 e^{-\theta \Gamma_{48}}
  \left(\sin^2\frac{\alpha}{2} B_1 - \cos^2\frac{\alpha}{2} B_2 \right)
-e^{\theta \hat\gamma \Gamma_6}
  \left(\cos^2\frac{\alpha}{2} B_1 - \sin^2\frac{\alpha}{2} B_2 \right)
\right\} \Gamma_{45}
\nn\\&\hskip2cm \left.
+\sin\frac{\alpha}{2}\cos\frac{\alpha}{2}
\left( e^{-\theta \Gamma_{48}} + e^{\theta \hat\gamma \Gamma_6}\right)
\left(
- \Gamma_{68} \Gamma_{11}
+B_1 \Gamma_{78}
-B_2 \Gamma_4 \Gamma_{11}
+B_1 B_2 \Gamma_5
\right)
\right] \epsilon_0 \,.
\end{align}
Since the coefficient of $\Gamma_{68} \Gamma_{11}$ term does not
involve $B_1$ nor $B_2$, this needs to be projected to either constant
or another gamma matrix structure.
However, it will not be compatible with the projection conditions
above, and then we conclude that there is no BPS solution with these
conditions.

   In summary: We cannot find the BPS configuration for D4 branes wrapping on $\mathbf{CP}^2$ or $\mathbf{CP}^1\times \mathbf{CP}^1$ with magnetic fields turned on.

\subsubsection{D6-brane}
\label{sec:D6}

We now consider a D6-brane wrapping on the whole $\mathbf{CP}^3$
and having a spike.
The ansatz we take is
\begin{center}
\begin{tabular}{r|cccc|cccccc}
 & $t$ & $x$ & $y$ & $r$ & $\alpha$ & $\chi$ & $\theta_1$ & $\vap_1$
 & $\theta_2$ & $\vap_2$  \\ \hline
D6 & $\sigma^0$ & $0$  & $0$ & $r(\alpha,\theta_1,\theta_2)$ & $\sigma^1$ & $\sigma^2$
 & $\sigma^3$ & $\sigma^4$ & $\sigma^5$  & $\sigma^6$
\end{tabular}
\end{center}
with $E_\alpha = \tilde{R}^{-2} \mathcal{F}_{\sigma^0 \sigma^1} \,,
\,  E_1 = \tilde{R}^{-2} \mathcal{F}_{\sigma^0 \sigma^3} \,,
\,  E_2 = \tilde{R}^{-2} \mathcal{F}_{\sigma^0 \sigma^5}$ turned on.
Then
\begin{align}
  \sqrt{- \det (g+\mathcal{F})} =&
\tilde{R}^7 \sin^2 \frac{\alpha}{2} \cos^2 \frac{\alpha}{2}
\sin\theta_1 \sin\theta_2 \sqrt{H} \,,
\\
H(r(\alpha)) = &
r^2 \left(1 + \frac{r_\alpha^2}{r^2} \right)
  \left(c_\alpha^2 + \frac{r_1^2}{r^2} \right)
  \left(s_\alpha^2 + \frac{r_2^2}{r^2} \right)
-E_\alpha^2
  \left(c_\alpha^2 + \frac{r_1^2}{r^2} \right)
  \left(s_\alpha^2 + \frac{r_2^2}{r^2} \right)
\nn\\&
-E_1^2
\left(1 + \frac{r_\alpha^2}{r^2} \right)
  \left(s_\alpha^2 + \frac{r_2^2}{r^2} \right)
-E_2^2
\left(1 + \frac{r_\alpha^2}{r^2} \right)
  \left(c_\alpha^2 + \frac{r_1^2}{r^2} \right)
 \,,
\end{align}
where $r_\alpha = \partial_\alpha r$, $r_{1,2} = \partial_{\theta_{1,2}}
r$ and $c_\alpha = \cos \alpha/2, s_\alpha = \sin \alpha/2$.

The projector is given by
\begin{align}
  \Gamma=&
\frac{1}{\sin\theta_1 \sin\theta_2 \sqrt{H}}
\Gamma_I \Gamma_{II} \,,
\nn\\
\Gamma_I =&
-r \Gamma_0
   \left(r\Gamma_4 - \frac{r_\alpha}{r}\Gamma_3 \right)
   \left(c_\alpha \Gamma_6 - \frac{r_1}{r}\Gamma_3 \right)
   \left(s_\alpha \Gamma_8 - \frac{r_2}{r}\Gamma_3 \right)
+E_\alpha
   \left(c_\alpha \Gamma_6 - \frac{r_1}{r}\Gamma_3 \right)
   \left(s_\alpha \Gamma_8 - \frac{r_2}{r}\Gamma_3 \right) \Gamma_{11}
\nn\\&
-E_1
   \left(r\Gamma_4 - \frac{r_\alpha}{r}\Gamma_3 \right)
   \left(s_\alpha \Gamma_8 - \frac{r_2}{r}\Gamma_3 \right) \Gamma_{11}
+E_2
   \left(r\Gamma_4 - \frac{r_\alpha}{r}\Gamma_3 \right)
   \left(c_\alpha \Gamma_6 - \frac{r_1}{r}\Gamma_3 \right) \Gamma_{11}
\,, \nn\\
\Gamma_{II}=&
\Gamma_5
\left(
\sin\theta_1 \Gamma_7 + \sin \frac{\alpha}{2}\cos\theta_1 \Gamma_5
 \right)
\left(
\sin\theta_2 \Gamma_9 - \cos \frac{\alpha}{2}\cos\theta_2 \Gamma_5
 \right) \,.
\end{align}
We then impose the following projection conditions:
\begin{align}
  \hat\gamma \Gamma_3 \epsilon_0 =& \epsilon_0 \,,
\\
 \Gamma_{03}\Gamma_{11} \epsilon_0 =& v \epsilon_0 \,,
\\
 \Gamma_{0456789} \epsilon_0 =& u \epsilon_0 \,,
\\
\hat\gamma \Gamma_{11} \epsilon_0 =& -\Gamma_{45} \epsilon_0
= \Gamma_{67} \epsilon_0 = -\Gamma_{89} \epsilon_0 \,,
\end{align}
and by using the first two conditions,
it is easy to see that $\mathcal{L} \epsilon_0 =
e^{v(-\xi_1 + \xi_2 -\xi_3 +\xi_4) \Gamma_0} \epsilon_0$
and since $\Gamma_0$ commutes with $\mathcal{K}$ and $\Gamma$,
$\mathcal{L}$ will decouple from the BPS condition.
$\mathcal{K} \epsilon_0$ is also simplified as
$  \mathcal{K}\epsilon_0 = e^{-\frac{\alpha}{2} \hat\gamma \Gamma_4} \epsilon_0
\equiv \mathcal{K}' \epsilon_0$.
Then
\begin{align}
  \Gamma_I \Gamma_{II} \mathcal{K}'
=& \mathcal{K}' \Gamma_{II}
\left[
-r\Gamma_0
  \left(\Gamma_4 - \frac{r_\alpha}{r}\Gamma_3 \right)
  \left(\frac{r_2}{r}c_\alpha \Gamma_{36} - \frac{r_1}{r} s_\alpha
\Gamma_{38} \right)
\right.\nn\\& \left.\hskip0.5cm
-\left\{
E_\alpha c_\alpha s_\alpha \Gamma_{68}
+E_\alpha \frac{r_1 r_2}{r^2}
+\left( \Gamma_4 - \frac{r_\alpha}{r}\Gamma_3 \right)
\left( \frac{r_2}{r}E_1 - \frac{r_1}{r}E_2 \right) \Gamma_3
\right\}\Gamma_{11}
\right.\nn\\&\left.
+\left\{
-r\Gamma_0
 \left( \Gamma_4 - \frac{r_\alpha}{r}\Gamma_3 \right)
 \left( c_\alpha s_\alpha \Gamma_{68} + \frac{r_1 r_2}{r^2} \right)
\right.\right.\nn\\&\left.\left.
-\left(-E_\alpha
    \left(\frac{r_2}{r}c_\alpha \Gamma_{63} + \frac{r_1}{r}s_\alpha\Gamma_{38} \right)
+\left( \Gamma_4 - \frac{r_\alpha}{r}\Gamma_3 \right)
 (-E_1 s_\alpha \Gamma_8 + E_2 c_\alpha \Gamma_6 )
\right)\Gamma_{11}
\right\} e^{-\alpha\hat\gamma \Gamma_4}
\right] \,.
\end{align}
We then consider the last $[ \cdots ]$ factor on $\epsilon_0$.
By applying the projection conditions, we have
\begin{align}
\big[ \cdots \big](\text{on}\, \, \epsilon_0) =&
 \Gamma_4
 \left(
 \left(
  v r_2 - E_2 \cos\alpha - \frac{r_\alpha}{r}\sin\alpha E_2
  +\frac{r_2}{r} E_\alpha \sin\alpha \right)c_\alpha \Gamma_6
\right.\nn\\&\left.\hskip1em
 -\left(
  v r_1 - E_1 \cos\alpha - \frac{r_\alpha}{r}\sin\alpha E_1
  +\frac{r_1}{r} E_\alpha \sin\alpha \right)s_\alpha \Gamma_8
 \right)\Gamma_{11}
\nn\\&
+\left(
-E_\alpha + v r_\alpha \cos\alpha - v r \sin\alpha
\right)s_\alpha c_\alpha \Gamma_{68}\Gamma_{11}
\nn\\&
+\frac{1}{r^2}
 \left(
 r_\alpha r_2 E_1 - r_\alpha r_1 E_2 - r_1 r_2 E_\alpha
 +v r_\alpha r_1 r_2 \cos\alpha - v r r_1 r_2 \sin\alpha
 \right) \Gamma_{11}
\nn\\&
+\left\{
 \left(
 -\frac{r_\alpha r_2}{r} - v \frac{r_2}{r}E_\alpha \cos\alpha
 + v \frac{r_\alpha}{r} E_2 \cos\alpha - v E_2 \sin\alpha
 \right) c_\alpha \Gamma_6
\right.\nn\\&\left.\hskip1em
- \left(
 -\frac{r_\alpha r_1}{r} - v \frac{r_1}{r}E_\alpha \cos\alpha
 + v \frac{r_\alpha}{r} E_1 \cos\alpha - v E_1 \sin\alpha
 \right) s_\alpha \Gamma_8
\right\}\Gamma_0
\nn\\&
+\left(
-v \frac{1}{r} (r_2 E_1 - r_1 E_2) - \frac{r_1 r_2}{r} \cos\alpha
-\frac{r_\alpha r_1 r_2}{r^2}\sin\alpha
\right) \Gamma_{04}
\nn\\&
-(r\cos\alpha + r_\alpha \sin\alpha) s_\alpha c_\alpha \Gamma_{0468}
\,.
\end{align}
Since $\Gamma_{II}$ are terms of $\Gamma_5, \Gamma_7$ and $\Gamma_9$
and will not become terms with $\Gamma_{11}$ after projection
conditions, we impose here all the terms proportional to
$\Gamma_{11}$ to vanish and have
\begin{align}
  E_\alpha =& v (r_\alpha \cos\alpha - r\sin\alpha )
\,, \qquad
 E_1 = v r_1 \cos\alpha \,, \qquad
 E_2 = v r_2 \cos\alpha \,.
\end{align}
By plugging these solutions to the BPS equation again, we get
\begin{align}
&  \epsilon_0 =
\frac{1}{\sin\theta_1 \sin\theta_2 \sqrt{H}}
\nn\\&\times
\left[ \left(
\frac{r_\alpha r_2}{r} c_\alpha^2 \sin\theta_1 \cos\theta_2
+\frac{r_\alpha r_1}{r} s_\alpha^2 \cos\theta_1 \sin\theta_2
\right.\right.\nn\\&\left.\hskip2em
+(r\cos\alpha + r_\alpha \sin\alpha)
 \left(
 -\frac{r_1 r_2}{r^2} c_\alpha s_\alpha \cos\theta_1 \cos\theta_2
 -c_\alpha s_\alpha \sin\theta_1 \sin\theta_2
 \right)
\right) \mathbf{1}
\nn\\&+
\left(
-\frac{r_\alpha r_2}{r} c_\alpha^2 s_\alpha \cos\theta_1 \cos\theta_2
+\frac{r_\alpha r_1}{r} s_\alpha \sin\theta_1 \sin\theta_2
\right.\nn\\&\left.\hskip2em
+(r\cos\alpha + r_\alpha \sin\alpha)
 \left(
 -\frac{r_1 r_2}{r^2} c_\alpha \sin\theta_1 \cos\theta_2
 +c_\alpha s_\alpha^2 \cos\theta_1 \sin\theta_2
 \right)
\right) \Gamma_{46}
\nn\\&+
\left(
\frac{r_\alpha r_2}{r} c_\alpha \sin\theta_1 \sin\theta_2
+\frac{r_\alpha r_1}{r} s_\alpha^2 c_\alpha \cos\theta_1 \cos\theta_2
\right.\nn\\&\left.\hskip2em
+(r\cos\alpha + r_\alpha \sin\alpha)
 \left(
 \frac{r_1 r_2}{r^2} s_\alpha \cos\theta_1 \sin\theta_2
 +c_\alpha^2 s_\alpha \sin\theta_1 \cos\theta_2
 \right)
\right) \Gamma_{48}
\nn\\&+
\left(
-\frac{r_\alpha r_2}{r} c_\alpha s_\alpha \cos\theta_1 \sin\theta_2
+\frac{r_\alpha r_1}{r} s_\alpha c_\alpha \sin\theta_1 \cos\theta_2
\right.\nn\\&\left.\hskip2em \left.
+(r\cos\alpha + r_\alpha \sin\alpha)
 \left(
 \frac{r_1 r_2}{r^2} \sin\theta_1 \sin\theta_2
 -c_\alpha^2 s_\alpha^2 \cos\theta_1 \cos\theta_2
 \right)
\right) \Gamma_{68}
\right] \epsilon_0 \,.
\end{align}
The resulting $\Gamma$ matrices are not commuting with the projection
conditions and thus all the coefficients need to vanish.
This condition can be solved by
\begin{align}
  \frac{r_\alpha r_1}{r} =&
-(r\cos\alpha + r_\alpha \sin\alpha )
 c_\alpha s_\alpha \frac{\cos\theta_1}{\sin\theta_1} \,,
\\
  \frac{r_\alpha r_2}{r} =&
-(r\cos\alpha + r_\alpha \sin\alpha )
 c_\alpha s_\alpha \frac{\cos\theta_2}{\sin\theta_2} \,,
\\
  \frac{r_1 r_2}{r^2} =&
 c_\alpha^2 s_\alpha^2\frac{\cos\theta_1}{\sin\theta_1}
 \frac{\cos\theta_2}{\sin\theta_2} \,.
\end{align}
Therefore the BPS equation is now
\begin{align}
  \frac{v}{\sqrt{H}}A \epsilon_0 =& \epsilon_0 \,,
\end{align}
and
\begin{align}
  A=& \frac{\text{coefficient of $\mathbf{1}$ term} }{\sin\theta_1
    \sin\theta_2}
\nn\\=&
-(r\cos\alpha + r_\alpha \sin\alpha ) c_\alpha s_\alpha
\left(1+s_\alpha^2\frac{\cos^2 \theta_1}{\sin^2 \theta_1}  \right)
\left(1+c_\alpha^2\frac{\cos^2 \theta_2}{\sin^2 \theta_2}  \right) \,.
\\
H=& (r\cos\alpha + r_\alpha \sin\alpha )^2 c_\alpha^2 s_\alpha^2
\left(
1+
\left(1-\cos^2\alpha \frac{r^2+r_\alpha^2}{r_\alpha^2} \right)
\left(
s_\alpha^2 \frac{\cos^2 \theta_1}{\sin^2 \theta_1}
+c_\alpha^2\frac{\cos^2 \theta_2}{\sin^2 \theta_2}
\right)
\right.\nn\\&\left.
+\left(1-2\cos^2\alpha \frac{r^2+r_\alpha^2}{r_\alpha^2} \right)
c_\alpha^2 s_\alpha^2 \frac{\cos^2 \theta_1}{\sin^2 \theta_1}
\frac{\cos^2 \theta_2}{\sin^2 \theta_2}
\right) \,.
\end{align}
If $H=A^2$, there will be a solution for $v=-1$.
However, it does not seem to be the case.


 \end{document}